\documentclass[%
 reprint,
 nofootinbib,
 amsmath,amssymb,
 aps,
 pra,
 floatfix,
 twocolumn
]{revtex4-1}

\pdfoutput=1
\usepackage[spanish,english]{babel}
\usepackage[utf8]{inputenc}
\usepackage[T1]{fontenc}
\newcommand{\ignore}[1]{}
\usepackage{fancyhdr}
\usepackage{xcolor}
\usepackage{pgf}
\usepackage{graphicx}
\usepackage{appendix}
\usepackage{mathtools}
\usepackage{makecell}
\usepackage{multirow}
\usepackage[compact]{titlesec} 
\usepackage{physics}

\newcommand{\toronto}{\texttt{toronto }}
\newcommand{\cnot}{\textsc{cnot} }

\begin{document}

\title{Benchmarking Characterization Methods for Noisy Quantum Circuits}
\date{23 DEC 2021}
\author{Megan L. Dahlhauser}
\email{mldahlh@sandia.gov}
\homepage{https://orcid.org/0000-0002-5747-9695; now at Sandia National Laboratories}
\affiliation{Quantum Science Center, Oak Ridge National Laboratory, Oak Ridge, Tennessee, USA}
\affiliation{Bredesen Center for Interdisciplinary Research and Graduate Education, University of Tennessee, Knoxville, USA}
\author{Travis S. Humble}
\email{humblets@ornl.gov}
\homepage{https://orcid.org/0000-0002-9449-0498}
\thanks{This manuscript has been authored by UT-Battelle, LLC under Contract No. DE-AC05-00OR22725 with the U.S. Department of Energy. The United States Government retains and the publisher, by accepting the article for publication, acknowledges that the United States Government retains a non-exclusive, paid-up, irrevocable, world-wide license to publish or reproduce the published form of this manuscript, or allow others to do so, for United States Government purposes. The Department of Energy will provide public access to these results of federally sponsored research in accordance with the DOE Public Access Plan. (http://energy.gov/downloads/doe-public-access-plan).}
\affiliation{Quantum Science Center, Oak Ridge National Laboratory, Oak Ridge, Tennessee, USA}
\affiliation{Bredesen Center for Interdisciplinary Research and Graduate Education, University of Tennessee, Knoxville, USA}
\begin{abstract}
Effective methods for characterizing the noise in quantum computing devices are essential for programming and debugging circuit performance. Existing approaches  vary in the information obtained as well as the amount of quantum and classical resources required, with more information generally requiring more resources. Here we benchmark the characterization methods of gate set tomography, Pauli channel noise reconstruction, and empirical direct characterization for developing models that describe noisy quantum circuit performance on a 27-qubit superconducting transmon device. We evaluate these models by comparing the accuracy of noisy circuit simulations with the corresponding experimental observations. We find that the agreement of noise model to experiment does not correlate with the information gained by characterization and that the underlying circuit strongly influences the best choice of characterization approach. Empirical direct characterization scales best of the methods we tested and produced the most accurate characterizations across our benchmarks.

\end{abstract}
\maketitle
\section{Introduction}
\label{sec:intro}
\par
Quantum computers are a promising technology anticipated to provide boosts to many  computational workflows \cite{watersim2020, kandala2017hardware, qSVM2015}, but the presence of noise in current quantum processing units (QPUs) leads to unexpected behaviors and performance of quantum circuits \cite{Tannu_2019, rudinger2019probing}.  Characterizing how noise present in these devices impacts a programmed circuit is an important part of assessing progress toward practical quantum computing and ultimately quantum computational advantage \cite{arute2019quantum,zhong2020quantum}.
\par
Several different methods for characterizing QPU have been proposed with these methods varying in the type of characterization tests used and the type of information received. In particular, most approaches to characterization appear on a spectrum of information gain versus scalability--in general the more information a protocol can provide, the less scalable the method, and vice versa \cite{Nielsen_2020}. On one end of this spectrum are coarse-grained characterization methods that are scalable but provide a limited amount information about the underlying physical noise process \cite{dahlhauserPRA}. On the other end of this spectrum are methods like gate set tomography (GST), that provide a finer grained model about the individual noise processes observed in experiment but are resource-intensive to perform \cite{gstnature}. In between these extremes are methods such as Pauli channel noise reconstruction (NR) that yield a modest amount of physics detail but remain limited in the description of the noise itself \cite{learningquantumnoise2020, estimationPaulichannels2020}.
\par
Our goal is to evaluate the performance of these quantum computing characterization protocols. We test the performance of GST, NR, and EDC on a variety of components and contexts. These protocols are selected because their motivations are different, as are their advantages, disadvantages, and resource consumption, but their outputs are complementary. They have commonalities which we use in developing comprehensive comparisons among these protocols. In particular, they utilize the description of Pauli noise, as described in Section \ref{sec:background}. We use this common language to design tests which identify the effectiveness of the protocols at characterizing quantum computers.
\par
Here we test these three characterization approaches side-by-side. We use EDC, NR, and GST to develop noise models which characterize a 27-qubit quantum computer in experiment. These methods are introduced in Section \ref{sec:background}. In Section \ref{sec:methods} we outline our method to compare the accuracy of the characterizations by using the noise models in simulation and comparing to experiment. We also detail our experiments to execute each characterization protocol, the quantum computers used in experiment, and the simulation methods. We present our results in Section \ref{sec:results}, including quantum and classical resource costs and the characterization results and accuracy compared to experiment. We offer final conclusions in Section \ref{sec:conc}.
\section{Background}
\label{sec:background}
\subsection{Empirical Direct Characterization}
\label{subsec:edc}
Empirical direct characterization (EDC) is a method for generating effective models of noisy quantum circuits using experimental characterizations. This method is introduced in reference \cite{dahlhauserPRA}. EDC is based on modeling quantum circuits using a suite of test circuits which construct a set of noisy subcircuit models. We compose the subcircuit models to generate noise models for larger and more complex quantum circuits. A key element in EDC is testing the fidelity of these circuit noise models against experimental data such that iterative refinements to the test circuits and noise models can be made. 
\par
Empirical direct characterization yields a coarse-grained noise model which can be tailored via dynamic tuning to every experiment execution and iterative refinements to best suit the characterization task \cite{dahlhauserPRA}. EDC scales linearly with the number of components, e.g.~qubits and gates. We summarize the complete procedure as follows.
\begin{enumerate}
    \item Identify ideal circuit $C$.
    \item \label{procedure:decomposition} Decompose the circuit into set $S(C)=\{S_i\}$ of ideal subcircuits $S_i$.
    \item \label{procedure:testcircuits} Select set of test circuits $T=\{T_i\}$ which define an input state and ideal outcome for each element in $S$.
    \item \label{procedure:noisemodel} Propose a noisy subcircuit model $M_i=M(S_i,p_i)$ for each element in $S$ parameterized by $p_i$.
    \item Implement and execute $T$ on QPU to generate experimental characterizations $H_i=(T_i,R_i)$ using results $R_i$ returned from QPU.
    \item Using set of characterizations $H=\{H_i\}$, fit noise parameters $p_i$ based on calculated expected probabilities for each $M_i$.
    \item \label{procedure:TVD} Compose the noisy circuit model $M(C,p)$ for the target circuit and compare the actual executed circuit $A=(C,R_C)$ with recorded results $R_C$ from the QPU to the noisy circuit model using $d_{TV}(A,M)$.
    \item \label{procedure:refine} If $d_{TV}$ is not at threshold return to \ref{procedure:decomposition}, apply refinements to \ref{procedure:decomposition}, \ref{procedure:testcircuits}, and \ref{procedure:noisemodel}, and continue to \ref{procedure:TVD} until threshold is met.
\end{enumerate}

\par
For step \ref{procedure:refine}, refinements to step \ref{procedure:decomposition} include additional elements selected from the set $g$, addition of compositions of elements in $g$ such that the test components are larger, or addition of elements to $g$ not explicitly represented in $\mathcal{G}$. Refinements to step \ref{procedure:testcircuits} include additional initializations as test circuits. Refinements to step \ref{procedure:noisemodel} include additional noise model parameters $p_i$ or different noise channels to define $M$.
\subsection{Gate Set Tomography}
\label{subsec:gstintro}
Gate set tomography (GST) is a method for extracting quantitative and qualitative information about quantum gates implemented in a quantum computer \cite{gstnature,greenbaum2015introduction}. It arose as an extension of quantum process tomography (QPT) \cite{ibm_gst_2013,blumekohout2013robust}. 
\par
Quantum process tomography characterizes a quantum gate by generating an estimate of the process matrix or the Pauli transfer matrix by measuring the components of a quantum gate operating on a prepared quantum state. The QPT protocol assumes that the quantum state preparation and the measurement are either known or error-free. However, this is generally not the case in experiment, because state preparation and measurement (SPAM) errors are prevalent in many, if not all quantum processing units (QPUs) to date. Furthermore, in practice SPAM errors can often be the result of QPU components that QPT would be used to characterize. Because of this, QPT can be inaccurate in realistic quantum computing experiments. In particular, QPT can actually become less accurate as the gates improve \cite{ibm_gst_2013}. 
\par
Gate set tomography rectifies this self-consistency problem by defining and characterizing a set of gates that represents both the quantum gates of interest and the imperfect state preparation and measurement operations. By characterizing the full set of gates at once, GST is able to more accurately estimate the true quantum gates because SPAM operations are characterized explicitly.
\par
Despite requiring more quantum experiments to gather the necessary information to perform GST than quantum process tomography, the lessened sensitivity to SPAM errors is expected to be vital for understanding how to utilize quantum error correction on near-term devices. The degradation of QPT gate characterization results due to the influence of SPAM can be highly problematic. This is particularly true for determining fault-tolerance thresholds, which have stricter conditions on gate error than on SPAM error. Quantum process tomography is unlikely to give accurate threshold estimates when SPAM error is highly prevalent compared to gate error \cite{greenbaum2015introduction}.
\par
Gate set tomography completely characterizes
\begin{equation}
\mathcal{G} = \{ |\rho\rangle\rangle, \langle\langle E|, G_0,...,G_k \}
\end{equation}
where $|\rho\rangle\rangle$ represents the initial state, $\langle\langle E|$ is a POVM, and each $G_k$ is a quantum gate. The set $\mathcal{F}=\{F_1,...,F_N\}$ is defined as the SPAM gates which operate as $|\rho_j\rangle\rangle=F_j|\rho\rangle\rangle$ and $\langle\langle E_i|=\langle\langle E|F_i$. Every $F_n$ must be composed of gates from gate set $\mathcal{G}$; therefore the set $\mathcal{G}$ must include gates sufficient to compose the full set of states and measurements. One example of such a gate set is $\mathcal{G} = \{\{\},X_{\pi/2},Y_{\pi/2},X_{\pi}\}$ with $\mathcal{F}=\mathcal{G}$ which includes the empty gate \{\}. Each gate $G_k$ can be reconstructed by measuring $\langle\langle E_i|G_k|\rho_j\rangle\rangle$. The GST protocol will characterize the full set $\mathcal{G}$ at once and only requires one initial state $\rho$ and one final measurement $E$. 
\par
The GST algorithm for one qubit is as follows \cite{greenbaum2015introduction}:
\begin{enumerate}
    \item Initialize to state $|\rho\rangle\rangle$
    \item For some $i,j,k$ of $i,j\in \{1...N\},k\in \{0...K\}$ apply gate sequence $F_i \circ G_k \circ F_j$
    \item Measure POVM $E$ which must be a positive semidefinite Hermitian operator with $I-E$ also positive semidefinite 
    \item Repeat steps 1-3 a large number of times n and per execution $r$ record $n_r=1$ if measurement is success or $n_r=0$ if failure
    \item Average the results of step 4 to get $m_{ijk}=\sum^n_{r=1} \frac{n_r}{n}$ which is a measurement of expectation value $p_{ijk}=\langle\langle E|F_iG_kF_j|\rho\rangle\rangle$
    \item Repeat steps 1-5 for all $i,j,k$
    \item Optional for additional independent measurements: repeat steps 1-5 to measure expectation values $p_i=\langle\langle E|F_i|\rho\rangle\rangle$
\end{enumerate}
\subsection{Randomized Compiling}
\label{subsec:RC}
\par
Randomized compiling (RC) is a method of transforming quantum circuits into a set of logically equivalent circuits by utilizing randomly selected twirling operators \cite{wallman2015noise}. First, a quantum circuit is expressed in cycles, which are each a single time step of parallelized quantum operators within the circuit with no more than one operation per qubit. These cycles are decomposed into ``easy'' gates which are assumed to have low or negligible error rates and ``hard'' gates which are assumed to have high error rates. Twirling operators are then injected around the hard gates which have the effect of tailoring the noise in the system to a stochastic Pauli channel. The injected twirling gates must be easy gates, and these are compiled together with the other easy gates in the cycle such that they become a single round of easy gates. 
\par
Randomized compiling can be used with a variety of twirling methods. We use Pauli twirling, which is one of the most commonly used twirling techniques. Pauli twirling is a method which turns a quantum operator into a Pauli channel,
\begin{equation}
    T_P(\varepsilon(\rho)) = \frac{1}{|\textrm{P}|}\sum_{P\in \textrm{P}} P\varepsilon(\rho) P^\dag = \sum_{P\in \textrm{P}}c_P P\rho P^\dag
\label{eq:paulitwirling}
\end{equation}
where $P$ is a Pauli matrix from the set \textrm{P} and the coefficients $c_P$ define the probability distribution over the Pauli operators. The set \textrm{P} is defined as the Pauli matrices $P^{\otimes n}$ for $n$ number of qubits in the system. The sampling set therefore grows exponentially in the register size, so for systems with large $n$ we may instead use randomized twirling, 
\begin{equation}
    T_P(\varepsilon(\rho)) = \frac{1}{N}\sum^{N}_{n=1}P_n\rho P_n^\dag
\end{equation}
where we select a limit of $N$ operators from which to sample. In the limit of the highest possible value of $N=4^n$, randomized Pauli twirling becomes equal to Eq. (\ref{eq:paulitwirling}) \cite{Cai2019}.
\par
\begin{figure}
    \centering
    \includegraphics[width=0.3\textwidth]{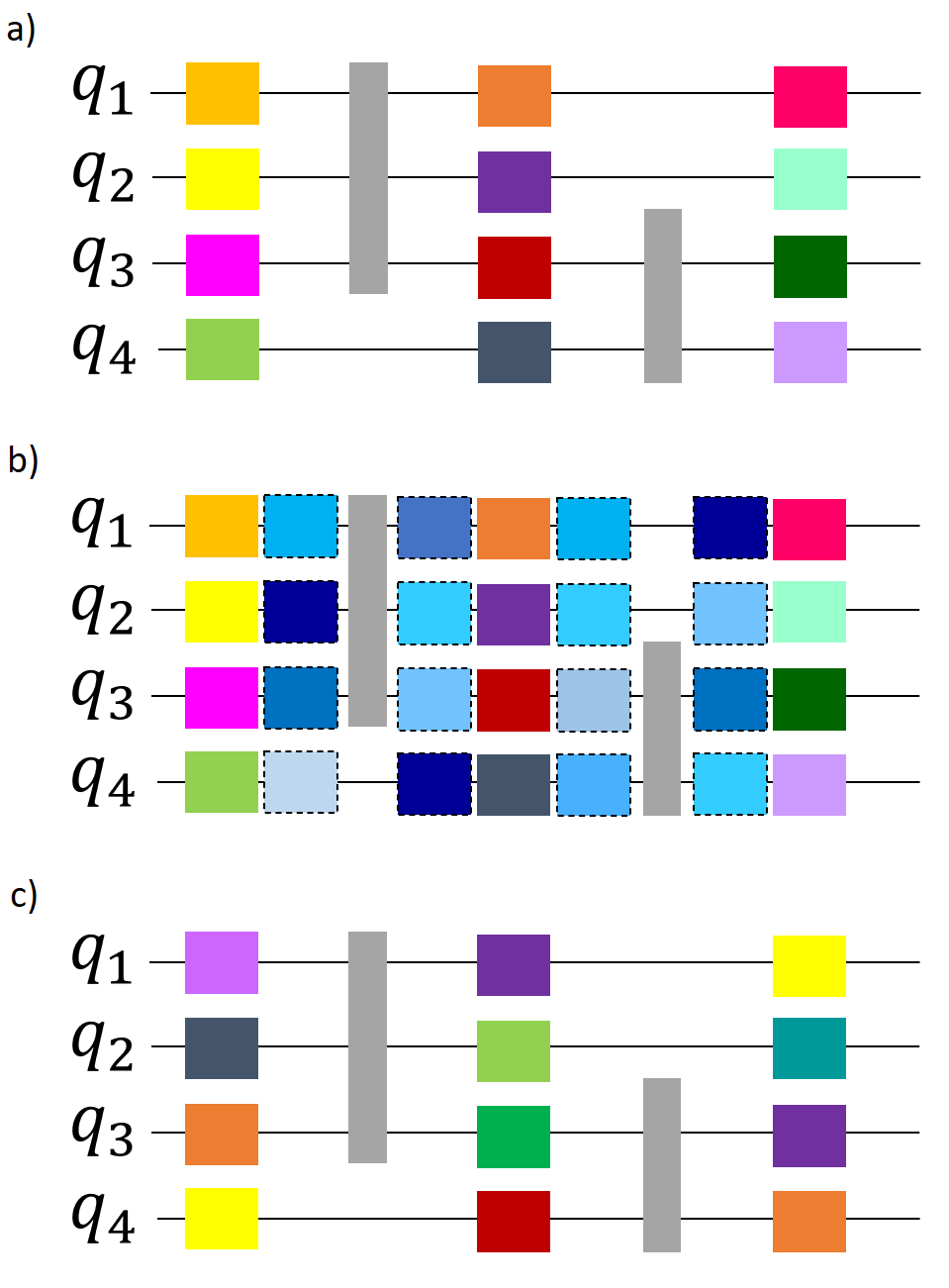}
    \caption{Graphical representation of randomly compiling a quantum circuit \cite{wallman2015noise,true-q}. Colored boxes represent easy gates; grey multi-qubit gates are considered to be hard gates in this example. Starting with a quantum circuit in a) we inject twirling gates in b) which are depicted by blue squares with dashed lines. Then in c) these gates are compiled together to form a randomly compiled circuit. This process is repeated for a set of randomly selected twirling gates to generate a set of $n$ twirled quantum circuits which together represent the randomly compiled quantum circuit.}
    \label{fig:rcdiagram}
\end{figure}

Randomized compiling is implemented by adding gates from the twirling group, which in our case are any Pauli gates from Eq.~\ref{eq:paulis} and the corresponding correction operator such that the overall unitary of the circuit is preserved. These added gates are compiled with neighboring easy gates which reduces the impact of randomized compiling on the circuit depth. This process is shown in Fig.~\ref{fig:rcdiagram}. The final output of randomized compiling is a set of quantum circuits with randomly applied operators. The results of a randomly compiled quantum circuit are taken as the sum of the results over the set of these circuits.
\begin{equation}
    X = \begin{pmatrix}
    0 & 1 \\ 1 & 0
    \end{pmatrix}, \ Y =
    \begin{pmatrix}
    0 & -i \\ i & 0
    \end{pmatrix}, \ Z = 
    \begin{pmatrix}
    1 & 0 \\ 0 & -1
    \end{pmatrix}
    \label{eq:paulis}
\end{equation}
\par
Pauli twirling has been used in several different contexts in quantum computing, from experiment reduction in characterization protocols to enhancement of computer performance \cite{Cai2019}. In randomized compiling, its purpose is to average the errors in the gate implementations into a stochastic Pauli channel. This has several benefits. Stochastic Pauli channels are more predictable and stable than other types of error such as coherent errors or spatial correlations among quantum components. By averaging the effects of these types of errors into a stochastic Pauli noise channel, we can estimate a description of the noise that is less complex than that of the uncompiled circuit. Randomized compiling is also expected to suppress error overall in the final results of compiled quantum circuits, at least in certain error regimes. For instance, average gate error is reduced in the case of over-rotation noise per gate with a factor of $10^{-2}$ difference in infidelity between easy and hard gates \cite{wallman2015noise}.
\subsection{Noise Reconstruction}
\label{subsec:NRoverview}
\par
Noise reconstruction (NR) is a protocol which enables estimation of process fidelities along with the associated error probabilities. It stems from the relationship between Pauli fidelities $f_i$ which measure how susceptible the Pauli operator $P_i$ is to noise,
\begin{equation}
    f_i=2^{-n}\text{Tr}(P_i\varepsilon(P_i))
\end{equation}
and the Pauli channel expression of Pauli error rates $p_i$ which express the likelihood of the occurrence of a Pauli operator $P_i$ as an error on state $\rho$,
\begin{equation}
    \varepsilon(\rho)=\sum_i p_i P_i\rho P_i
\end{equation}
These two metrics--Pauli fidelities and Pauli error rates--are related via the Walsh-Hadamard transform
\begin{equation}
    \boldsymbol{f_G}=\boldsymbol{W_{G,\text{P}^n}p}
\end{equation}
for a group of Paulis $G$, where $\boldsymbol{f_G}$ is the vector representation of the fidelities $f_g$ for elements $g \in G$ and $\boldsymbol{p}$ is the vector of Pauli error rates. The transform $\boldsymbol{W_{G,\text{P}^n}}$ maps from group $\text{P}^n$ to $G$, where $\text{P}^n$ is the quotient group of Paulis with its center. The columns of this transform that correspond to Paulis that differ only by an element that commutes with all $g$ are interchangeable, and therefore cannot produce the necessary reconstruction between fidelities and probabilities. Instead, we have to restrict the transform to the anticommutant of the group $G$, such that 
\begin{equation}
    \boldsymbol{f_G}=\boldsymbol{W_{G,\text{A}_G}p_{\text{A}_G}}
\end{equation}
In practice, applying the inverse Walsh-Hadamard transform to the fidelity vector can yield the corresponding error probabilities \cite{flammia2020efficient}.
\par
The NR algorithm is described below \cite{harper2019efficient}.
\begin{enumerate}
    \item Choose one- or two-qubit twirling sequences from the Clifford group (Hadamard, phase, and/or CNOT gates)
    \item Sample empirically to estimate the probability distribution from measurement outcomes
    \item Calculate the Walsh-Hadamard transform of this probability distribution
    \item Fit these transformed values to the exponential decay $Af^m$ dependent on sequence length $m$, yielding the fidelities $f$
    \item Perform reverse transform and project onto probability vector, which will reconstruct the entire list of effective qubit error rates
\end{enumerate}
This procedure converges to the estimate of the average noise \cite{flammia2020efficient}. It scales polynomially in the number of qubits and the number of error rates. But since the possible correlations depends on the number of qubits, the number of error rates scales exponentially in the number of qubits. To limit this scaling to polynomial rather than exponential, error correlations are limited in range according to the physically-motivated constraints of error correlations between a qubit and only its nearest neighbors as defined by the topology of the qubit register.
\section{Methods}
\label{sec:methods}
\par
We design several tests for metrics of interest. The primary metric is the accuracy of each method in capturing the fundamental behavior of the device. We evaluate this in two ways. First, we calculate the distance between the empirical results and results estimated using the selected protocols using noisy simulation with noise models parameterized by the characterization results. We use the total variation distance (TVD) defined in Eq.~\ref{eq:TVD} as the metric for this calculation. Second, we evaluate the ability of the protocols to predict performance of a QPU on a benchmark application. To do this, we identify quantum circuit implementations which are composed of components we have characterized. We gather experimental data for these applications and simulation data under noise models designed from characterization information from each protocol. We compare simulated results to empirical results and evaluate how close our simulation is to experiment using TVD. 
\begin{equation}
    d_{\textsc{tv}}(H_i, M_i) = \frac{1}{2}\sum_k{\Big \lvert r^{(H_i)}(k) - r^{(M_i)}(k) \Big \rvert}
    \label{eq:TVD}
\end{equation}
\par
Another metric of interest is efficiency, specifically how these protocols scale with the size of the quantum register. The scalability is often based on the dependencies of the algorithm, and the number of quantum experiments needed for each of our selected protocols to characterize a particular gate set on a selected qubit register is known. However, more precisely establishing the tradeoff between experiment count and accuracy of the characterization measured by TVD is a key metric for evaluating these methods. In particular, we measure the relationship between the experiment count of implementations of each protocol and the TVD between these characterizations used in noisy simulation and their associated empirical results. This relationship helps to identify thresholds for the achievable accuracy under a particular experiment count limit, for example, the practical limitation of maximum experiment count per job sent to a QPU.
\par
Classical processing and computing efficiencies are important considerations as well. For instance, classical computational resources are used in processing characterization data to generate protocol output. The efficiency of GST, NR, and EDC is dominated by the quantum computational resources rather than classical computational resources, but classical resource costs may be prohibitive for large quantum circuit simulations and optimization over large data sets, for example. 
\par
Our experiment design is outlined as follows.
\begin{enumerate}
    \item Select characterization protocols--GST, NR, EDC--which generate metrics such as process fidelity and error rates that predict low-level performance.
    \item Select a suite of test circuits to characterize. We use Bell-state preparation circuits and GHZ-state preparation circuits shown in Appendix \ref{app:bellghz}. 
    \item Select a suite of circuits to test the predictive capacity of each protocol's characterization output. We use the Bernstein-Vazirani algorithm, shown in Appendix \ref{app:bv}, implemented for all accessible secret string encodings.
    \item Select QPUs and collect experimental data for each protocol and each application circuit. We use the IBM Q suite of QPUs \cite{ibmqsuite}.
    \item Analyze characterization data to generate protocol output and noise models. This analysis includes calculating noise parameters that best fit the data and metrics such as process fidelities and noise rates per component, for example.
    \item Report on metrics of these results. This includes:
    \begin{itemize}
        \item Accuracy of noisy simulation based on measured characterization parameters in both application circuit performance and predicted performance in additional applications.
        \item Efficiency and scaling of methodology in computational resources, including time, quantum experiments and classical processing and analysis.
        \item Effectiveness of the translation of characterization data to a performance benchmark.
    \end{itemize}
\end{enumerate}
\subsection{Devices Tested}
\begin{figure}[ht]
    \centering
    \includegraphics[width=0.4\textwidth]{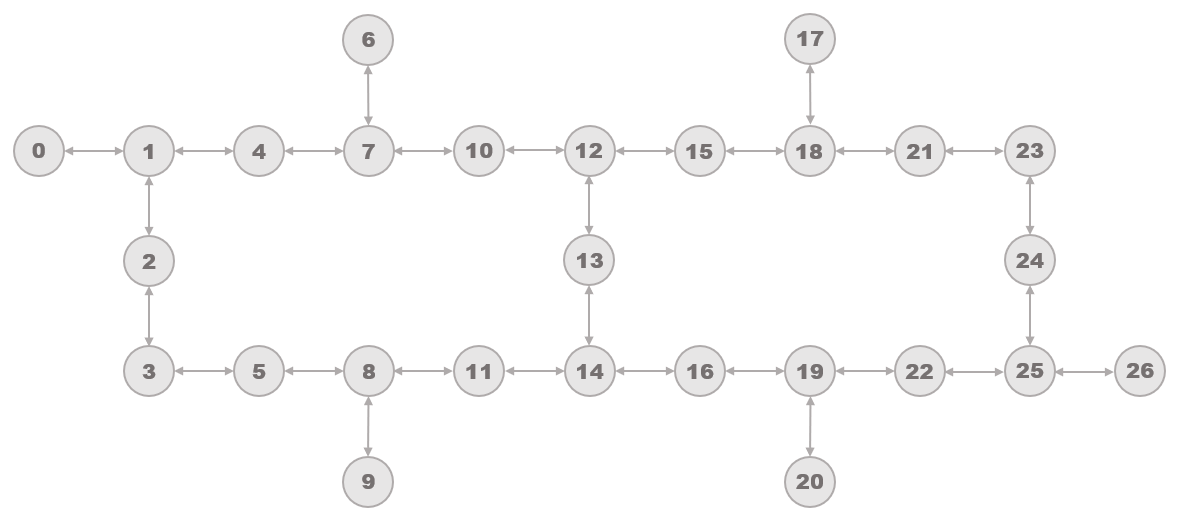}
    \caption{A graphical representation of the register layout of the 27-qubit \toronto QPU at the time of data collection. Each node corresponds to a register element and directional edges indicate the availability of a programmable two-qubit cross-resonance gate. }
    \label{fig:Tlayout}
\end{figure}

To gather empirical data for testing our benchmarking protocols we use the IBM Q suite of quantum processors (QPUs) \cite{ibmqsuite}. All of our selected characterization protocols may be straightforwardly executed on any QPU which has a gate-level interface, but we select the IBM suite because they are publicly available and provide an array of QPUs of differing register properties. We focus our experiments on \texttt{toronto}, a 27-qubit superconducting transmon device with layout as shown in Fig.~\ref{fig:Tlayout}, which has a limit of 900 circuits per job and the option to reserve dedicated time \cite{ibmqsuite}. The relatively large register size of \toronto compared to other QPUs available makes \toronto a good choice for testing the scalability of these protocols while also remaining well within the limits of classical simulation of quantum computers. The importance of a high circuits-per-job limit and dedicated QPU time is to keep a high throughput, which prevents the introduction of drift in the system noise \cite{wilson2020justintime}. 
\par
We map GHZ-state preparation circuits onto \toronto as illustrated in Table \ref{tab:ghzmapping}. To prepare each GHZ state of size $n$ qubits, we sequentially apply the gates shown in the table for each size 2 to $n$. As an example, for a $3$-qubit GHZ state, we apply an $H$ gate to qubit 0, a \cnot gate between qubits 0 and 1, and a \cnot gate between qubits 1 and 2. This mapping is not unique nor is it optimized for any performance gains. The use of \textsc{SWAP} gates would enable less rigid adherence to the spatial topology of the device, but their addition would likely introduce additional noise sources so we restrict gate selections to the layout as shown in Fig.~\ref{fig:Tlayout}.
\par
For the Bernstein-Vazirani (BV) algorithm circuits defined in Section \ref{app:bv}, we select oracle qubit 25 and secret string encoded qubits 22, 24, and 26. We limit our BV algorithm implementation to a total of 4 qubits because this is the maximum number of qubits we may use without introducing \textsc{SWAP} gates. These qubits are selected because they have comparable or slightly lower error rates than other identically-connected 4-qubit groupings on \toronto as measured by IBM's routine calibration data \cite{ibmqsuite}.
\begin{table}[ht]
    \centering
    \caption{GHZ-state preparation circuit mapping onto \texttt{toronto}'s topology. For each GHZ size $n$, the preparation circuit is built by applying the gates of all sizes $[2,n]$ in series.}
    \begin{tabular}{|l|l|l|}
        \hline
        \textbf{GHZ Size} & \textbf{Gate Added} & \textbf{Qubit Added} \\
        \hline
        2 & \makecell[l]{H(0)\\\cnot (0,1)} & \makecell[l]{0 \\ 1} \\ \hline
        3 & \cnot (1,2) & 2  \\ \hline
        4 & \cnot (2,3) & 3 \\ \hline
        5 & \cnot (3,5) & 5 \\ \hline
        6 & \cnot (5,8) & 8 \\ \hline
        7 & \cnot (8,9) & 9 \\ \hline
        8 & \cnot (8,11) & 11 \\ \hline
        9 & \cnot (11,14) & 14 \\ \hline
        10 & \cnot (14,13) & 13 \\ \hline
        11 & \cnot (13,12) & 12 \\ \hline
        12 & \cnot (12,10) & 10 \\ \hline
        13 & \cnot (10,7) & 7 \\ \hline
        14 & \cnot (7,6) & 6 \\ \hline
        15 & \cnot (7,4) & 4 \\ \hline
        16 & \cnot (12,15) & 15 \\ \hline
        17 & \cnot (15,18) & 18 \\ \hline
        18 & \cnot (18,17) & 17 \\ \hline
        19 & \cnot (18,21) & 21 \\ \hline
        20 & \cnot (21,23) & 23 \\ \hline
        21 & \cnot (23,24) & 24 \\ \hline
        22 & \cnot (24,25) & 25 \\ \hline
        23 & \cnot (25,26) & 26 \\ \hline
        24 & \cnot (25,22) & 22 \\ \hline
        25 & \cnot (22,19) & 19 \\ \hline
        26 & \cnot (19,20) & 20 \\ \hline
        27 & \cnot (19,16) & 16 \\ \hline
    \end{tabular}
    \label{tab:ghzmapping}
\end{table}
\subsection{Empirical Direct Characterization}
\par
We utilize a set of quantum circuits for EDC characterization experiments as outlined in Section \ref{subsec:edc}. To characterize asymmetric readout, we use circuits of $X$ and $XX$ gates done in parallel and in isolation with one operation per circuit. We also use a blank circuit with no operations which will return a zero state in the absence of noise because IBM QPUs are initialized to the all-zero state. To characterize the error on \cnot gates we use a set of Bell-state preparation circuit tests which are applied to each qubit coupling of \toronto according to Fig.~\ref{fig:Tlayout}. We use the Bell state because it is a subcircuit of the GHZ state and therefore a good candidate to characterize the GHZ-state preparation circuits. We use EDC to characterize GHZ-state preparations of qubit register size 2-27 on \texttt{toronto}. 
\subsection{Noise Reconstruction}
\label{subsec:charexpstq}
\par
Pauli channel noise reconstruction characterizes noise of randomly compiled (RC) circuits. These characterizations rely on Pauli twirling and utilize a similar structure of experiment design for quantum circuit characterization. These are outlined in Section \ref{subsec:NRoverview}.
\par
For NR, we use the True-Q software to generate circuit collections for execution on IBM QPUs and to calculate the Pauli channel descriptions from NR \cite{true-q}. This software is developed by the company Quantum Benchmark. Noise reconstruction is referred to as $k$-body noise reconstruction (KNR) in True-Q, so we use KNR for clarity in reporting our results. $k$-body refers to the number of gates for which an error description is estimated. For instance, a cycle with three parallel gates could be defined with up to $k=3$. Then if $k=2$, Pauli channels would be estimated for every two-gate subset within the cycle. We have a software-enforced limit of 20 qubits for experiment design, circuit generation, and results in True-Q, so we limit our experiment design of KNR for GHZ cycles and the RC GHZ circuits to the first 20 qubits of the 27-qubit GHZ mapping we use on \texttt{toronto}. Our KNR protocols for BV cycles and RC BV circuits are executed on a 4-qubit subset on \toronto and therefore do not reach this limit.
\par 
We design experiments using KNR to characterize the components of the GHZ-state preparation and BV circuits. Specifically, we use KNR to characterize the Hadamard and \cnot gates for the qubits used in the $n$-qubit GHZ-state preparation as well as the Hadamard, \textsc{cnot}, and $X$ gates used in the BV algorithm circuits. The Pauli error rates estimated with KNR can then be used as input to noisy simulation, which we compare to experiment to evaluate the accuracy of the KNR characterization.
\par 
The experiments for KNR are defined in terms of cycles. Because cycles must be one time step of a circuit, i.e. only one round of parallel gates, we select two different types of cycles to characterize for GHZ. We use a per-gate cycle design which defines one cycle per gate of the GHZ circuit. In the GHZ circuit example, each gate is necessarily a separate time step, so this cycle selection is the most natural decomposition for the GHZ-state preparation circuits. This yields a total of 20 unique cycles for our 2-20-qubit GHZ-state preparation circuits. 
\par
For Bernstein-Vazirani algorithm circuits, each timestep of the circuit is defined as one cycle. Preparation of secret bitstring encodings uses three \cnot gates which are applied such that the control qubit corresponds to any encoded `1's in the bitstring. These \textsc{cnot}s are characterized as one cycle each. Every BV secret string encoding is preceded and succeeded by parallel Hadamard gates on every qubit and an $X$ gate on the oracle qubit. The Hadamards are characterized as one cycle together and the $X$ gate is characterized as one cycle.
\subsection{Gate Set Tomography}
\par
Because GST is prohibitively intensive for qubit registers beyond a couple of qubits \cite{gstnature,greenbaum2015introduction}, we will limit characterization with GST to 2 qubits. We can use GST to characterize a gate set which contains a collection of single-qubit and \cnot gates and use the results to generate a Pauli noise model from the process matrix. This data will represent a standard to which we can compare our other techniques, as GST should yield the most accurate picture of the noise present in the Bell-state preparation example. 
\par
To run the GST protocol, we use the python implementation called pyGSTi, which stands for Python Gate Set Tomography Implementation \cite{pygsti}. This implementation provides a software code framework for generating a circuit collection for execution on a QPU and data analysis of quantities of interest including average gate fidelity and estimated process matrices. pyGSTi is developed by a team based at Sandia National Laboratories. 
\par
For our GST experiments, we use a standard model within the pyGSTi framework which contains the gate set \{$R_X(\frac{\pi}{2})$,$R_Y(\frac{\pi}{2})$,$R_Z(\frac{\pi}{2})$,$I$,\textsc{cnot}\}. We perform the standard GST analysis on our data set (maximum likelihood gate set tomography, or MLGST). This process estimates the gate set that is the best fit to the experimental data by maximizing the log-likelihood with the gate set probabilities \cite{pygsti}.
\subsection{Noise Models}
\par 
For EDC, our estimated noise models include isotropic depolarizing two-qubit channels. This channel $\epsilon_{DP}$ is defined in terms of $p_{DP}$ such that 
\begin{equation}
    \epsilon_{DP}(\rho) = (1-p_{DP})I\rho I + \frac{p_{DP}}{3}(X\rho X + Y\rho Y + Z\rho Z)
    \label{eq:isodepol}
\end{equation}
where 
$\epsilon_{\textrm{DP}}^{j,k} = \epsilon_{\textrm{DP}}^{j} \otimes \epsilon_{\textrm{DP}}^{k}$ for qubits $j,k$. 
\par
For KNR, our estimated noise models include stochastic Pauli channels of one and two qubits. In all of our experiments, we consider only $k=1$ because almost all of our cycles are defined with just one gate based on the structure of the GHZ and BV circuits. Incorporating correlated errors among subsets of gates in the cycles with parallelized gates might enhance the detail of the final noise models, but using $k=1$ is a necessary first step for characterizing all our selected cycles and is most comparable to other methods.
\par
As defined from Eq.~\ref{eq:paulitwirling}, stochastic Pauli noise channels are of the form
\begin{equation}
    \epsilon_{SP}(\rho) = \sum_{P\in \mathbf{P_d}^{\otimes n}} c_P P\rho P^{\dag}
\end{equation}
where each $P$ is an $n$-qubit Pauli matrix with dimension $d=2$ for qubits. The KNR protocol provides estimates of a set of probabilities $c_p$ and Pauli matrices $P$ which describe the noise of a cycle.
\par
For both EDC and KNR noise models, we also estimate an asymmetric readout channel. The asymmetric readout channel is defined in terms of $p_0$ and $p_1$ which are the probability of a bit flip in the measurement of state 0 and state 1, respectively. The probability of a correct measurement then follows directly as $(1-p_0)$ and $(1-p_1)$, respectively.
\par
For GST, we use the Pauli transfer matrix (PTM) to describe the noise model of our quantum gates. The elements of the PTM are defined as
\begin{equation}
    \text{PTM}_{i,j} = \frac{1}{d} \text{Tr}\{P_i\Lambda (P_j)\}
    \label{eq:PTM}
\end{equation}
for dimension $d=2$, Pauli matrices $P$ and quantum operation $\Lambda$. This PTM represents the noisy gate and can be applied directly in simulations.
\par
Gate set tomography also estimates state preparation and measurement (SPAM) errors. For two-qubit tomography experiments these SPAM parameters provide estimates of error on each two-bit measurement output in a 4x4 matrix. The matrix elements represent the probabilities of measuring each classical two-bit outcome given an expected outcome.
\subsection{Simulation Methods}
\par
For our simulations, we use Qiskit Aer \cite{Qiskit}. Aer is a quantum circuit simulator which can simulate ideal or noisy quantum circuits with a variety of methods. For our simulations we use Aer's statevector simulator which simulates quantum circuits by applying operators to the statevector which describes the quantum state of the qubit register. It can simulate any of the gates and noise models that we use for our tests but the size of the computation scales exponentially in the size of the qubit register. Consequently, for our GHZ-state preparation circuits with register sizes around 20+ qubits we use the Aer statevector simulator on the IBM Q backend. This is a dedicated classical computing resource which is optimized for quantum circuit simulation such that large simulations can be completed more quickly than on a personal computer.
\par
We model the noise in quantum circuits as an ideal quantum operator followed by a noise operator which represents the noise associated with the ideal operator when applied in experiment. This is a common but not unique method to describe noise in quantum systems \cite{qiskitdocs}. The quantum error functions that are native to the Aer simulator methods utilize this expression of quantum noise. We define our noise models in the Aer framework to implement them in simulation.
\par
For simulations of the Bell-state preparation circuit using the GST estimated noise model, we use the pyGSTi simulation capability. pyGSTi supports quantum circuit simulation that uses the estimated model results calculated directly from the GST protocol. Because GST reports a more complex model of the characterized gate set than the other methods, simulating the Bell state directly in pyGSTi provides the most accurate translation of GST model results to circuit outcomes.
\par
Because GST simulations are limited to the two-qubit example, we do not simulate the GHZ or BV circuits using the GST model. For the Bell-state preparation circuit, our GST model defines a noisy \cnot gate and a noisy Hadamard gate which is defined as a decomposition into a rotation about $Y$ by $\pi$/2 and two rotations around $Z$ by $\pi$/2. The GST model also includes the state preparation and measurement error which maps the probability of every two-qubit input state to be observed as each two-qubit output state.
\par
The code and data used in these experiments can be found at the public repository \cite{publicrepo_qccb}.
\subsection{Application Testing}
\par
The outcome of any measured quantum circuit is a bitstring of zeroes and ones. To evaluate the distance between two distributions of bitstring outcomes, we use the total variation distance (TVD). The TVD is given by
\begin{equation}
    d_{\textsc{tv}}(H, M) = \frac{1}{2}\sum_k{\Big \lvert r^{(H)}(k) - r^{(M)}(k) \Big \rvert}
\end{equation}
for two distributions $H$ and $M$ with probability $r$ of state $k$, just as in Eq.~\ref{eq:TVD}. The probability $r$ is calculated by the number of times the state $k$ is returned divided by the total number of measurements which comprise the distribution.
\par
Error propagation in the TVD calculation is given by
\begin{equation}
    \delta \text{TVD} = \frac{1}{2}\sqrt{(\delta\alpha_i)^2 + (\delta\alpha_j)^2 + (\delta\beta_i)^2 + (\delta\beta_j)^2 + ...} 
\end{equation}
for states $\alpha$, $\beta$, $...$ of two distributions labelled $i$ and $j$. The error of each state is given by 
\begin{equation}
    \delta\alpha = \sqrt{\frac{p(1-p)}{N}}
\end{equation}
for probability $p$ of measuring the state out of $N$ total measurements.
\par
The Bernstein-Vazirani (BV) algorithm is our selected application test. The circuits which implement the BV algorithm utilize a gate set closely related to the GHZ-state preparation circuits. We use this algorithm as a benchmark of performance. The output of a BV circuit in the absence of noise is the encoded secret string, so we compare the accuracy of our noisy simulation in returning the encoded secret string to the accuracy obtained in experiment from the QPU. The accuracy is defined as the number of times the encoded secret string is observed out of the total shot count of the circuit. This provides a means to benchmark the noise models used in simulations--the closer the accuracy agrees with experiment, the more likely the noise model accurately describes the QPU. 

\section{Results}
\label{sec:results}
\par
We report results of characterization and performance testing using our selected methodologies as presented in Section \ref{sec:methods}. We executed GST, NR, and EDC protocols on \toronto over a 12-hour period of dedicated QPU time on February 14, 2021. We executed the GST circuits first. Next we ran KNR experiments for GHZ cycles followed by the RC GHZ circuits. Then we executed the circuits for KNR for the BV cycles followed by the RC BV circuits. Interspersed among these were multiple runs of EDC circuits. Uncompiled GHZ and BV circuits were included in the jobs that execute EDC circuits. We refer to these uncompiled circuits as bare circuits (BC).
\subsection{Quantum Resources Usage}
\label{subsec:quantumresources}
\par
A central feature of characterization methods is their resource use and scalability. In Table \ref{tab:quantumresources}, we summarize the resource requirements of our experiments, in particular the amount of time taken to acquire results and the size of the computational jobs. All quantum experiments are sent to IBM Q devices as jobs with a limit of 900 circuits per job. The number of shots per circuit on these devices is limited to 8192. Because data was taken during a 12-hour window of dedicated QPU time, there were no queue wait times for any experiments. We record the amount of time taken for an experiment set as the wall clock time from the creation of the first job containing experiments for the protocol to the completion of the last job containing experiments for the protocol. In the cases of KNR and RC experiments, the job creation and validation are parallelized by the True-Q software interface which substantially decreases the total time taken for these experiments compared to that of GST and EDC. 
\begingroup
\setlength{\tabcolsep}{12pt} 
\renewcommand{\arraystretch}{2} 
\begin{table*}[htb]
    \centering
    \caption{Quantum resources used in our selected protocols for all experiments executed. Sequence lengths for KNR are the number of times the cycle of interest is repeated in each experiment.}
    \begin{tabular}{|l|l|l|l|l|}
        \hline
        \textbf{Method} & \textbf{Details} & \textbf{Circuits} & \textbf{Shots} & \textbf{Time} \\
        \hline
        GST & \makecell[l]{2-qubit gate set \\ \{$R_X(\frac{\pi}{2})$,$R_Y(\frac{\pi}{2})$,\\$R_Z(\frac{\pi}{2})$,$I$,\cnot\}} & \makecell[l]{20094} & \makecell[l]{1024} & \makecell[l]{2.28 hours} \\ \hline
        KNR (GHZ) & \makecell[l]{Per-gate cycles (20) \\ Sequence lengths 4,12}  & \makecell[l]{10440}  & \makecell[l]{128} & 17 minutes \\ \hline
        KNR (GHZ) & \makecell[l]{Parallelized cycles (3) \\ Sequence lengths 4,12}  & \makecell[l]{1620}  & \makecell[l]{128} & 11 minutes \\ \hline
        RC GHZ & \makecell[l]{2-20-qubit GHZ \\ circuits compiled into \\ 32 RC circuits each} & \makecell[l]{608}  & \makecell[l]{128} & 12 minutes \\ \hline
        KNR (BV) & \makecell[l]{Time step cycles (5) \\ Sequence lengths 4,12} & \makecell[l]{1980}  & \makecell[l]{128} & 6 minutes \\ \hline
        RC BV & \makecell[l]{All 3-bit strings\\compiled into 32 RC \\ circuits each} & \makecell[l]{256}  & \makecell[l]{128} & 6 minutes \\ \hline
        EDC & \makecell[l]{27-qubit \\ characterizations} & \makecell[l]{205} & \makecell[l]{8192} & \multirow{4}{*}{11 minutes}\\
        \cline{1-4}
        BC GHZ  & \makecell[l]{2-27-qubit GHZ \\ circuits} & \makecell[l]{26} & \makecell[l]{8192} & \\
        \cline{1-4}
        BC BV & \makecell[l]{All 3-bit strings} & \makecell[l]{8} & \makecell[l]{8192} & \\
        \hline
    \end{tabular}
    \label{tab:quantumresources}
\end{table*}
\endgroup
\par
For GST, the 2-qubit 5-gate set we characterize is computationally expensive. The qubit count and gate count are the primary drivers of the total experiment count necessary to build a GST estimate. For instance, reducing this to just a single qubit example of the same gate set without \cnot would reduce the circuit count by ten times. We use a shot count of 1024 which is the default shot count setting and generally ensures sufficient statistics. 
\par
For KNR, the primary factors which determine computational expense are the sequence lengths and the number of cycles. There are 20 cycles needed to characterize every component of the GHZ circuits, so we utilized a minimum sequence count to keep resource costs manageable. We use sequence lengths of 4 and 12 because the error rates of \cnot gates tend to be high so the performance degrades after a short sequence of gates. The error bars on these estimates are consequently larger however, as a result of fewer data points to fit the decay curve over multiple sequence lengths. We use a shot count of 128 because these protocols calculate estimates based on the decay functions, so the sampling size of each individual data point may be reduced \cite{true-q}. This also helps to manage the resource cost.
\par
The EDC circuit count includes the circuits which we use to characterize readout errors which are applied for both EDC and KNR noise models. The EDC circuit count increases linearly with the number of qubits and the number of operators to characterize. We use the maximum shot count for these tests because there are few enough circuits that the resource cost is still low.
\subsection{Experimental Data}
\label{subsec:performancedata}
\par
\begin{figure}
    \centering
    \includegraphics[width=0.48\textwidth]{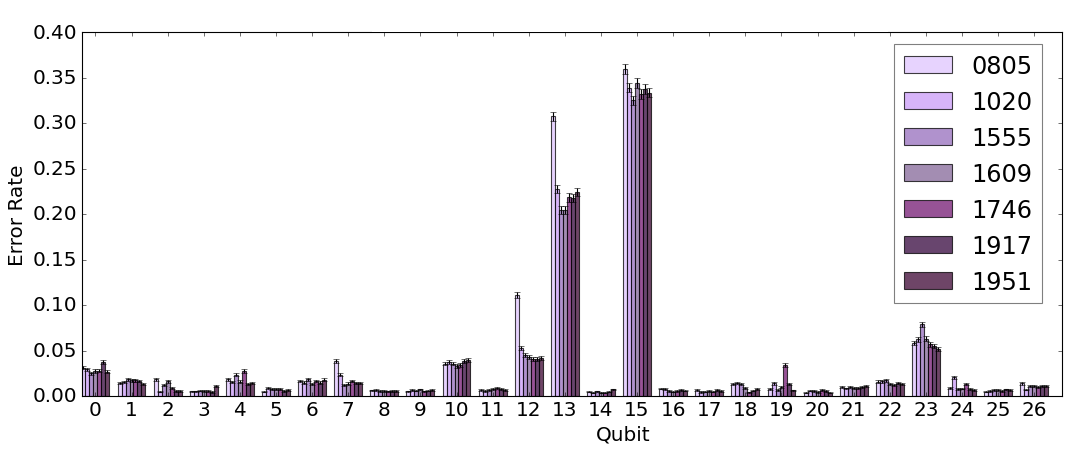}
    \caption{Error rate in readout of state 0 on \toronto using the EDC methodology. }
    \label{fig:p0compare}
\end{figure}
\begin{figure}
    \centering
    \includegraphics[width=0.48\textwidth]{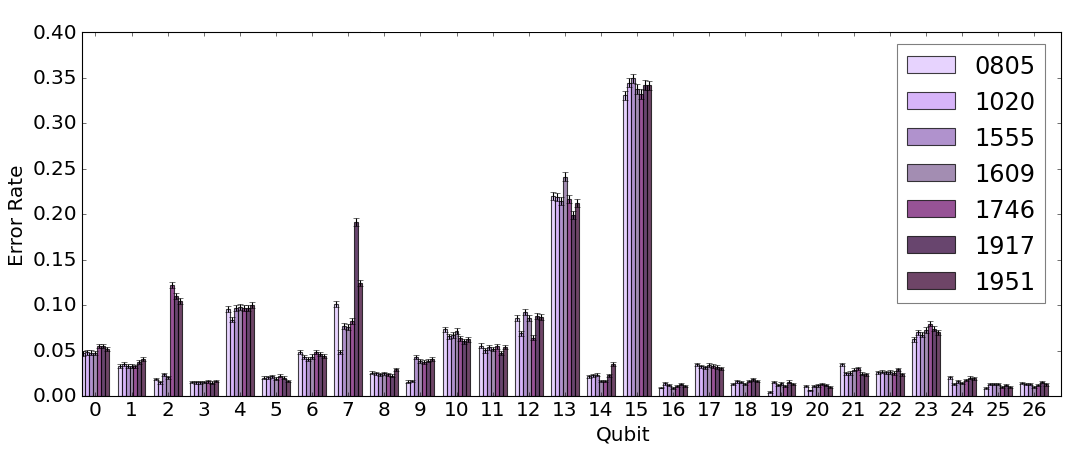}
    \caption{Error rate in readout of state 1 on \toronto using the EDC methodology.}
    \label{fig:p1compare}
\end{figure}
\begin{figure}
    \centering
    \includegraphics[width=0.48\textwidth]{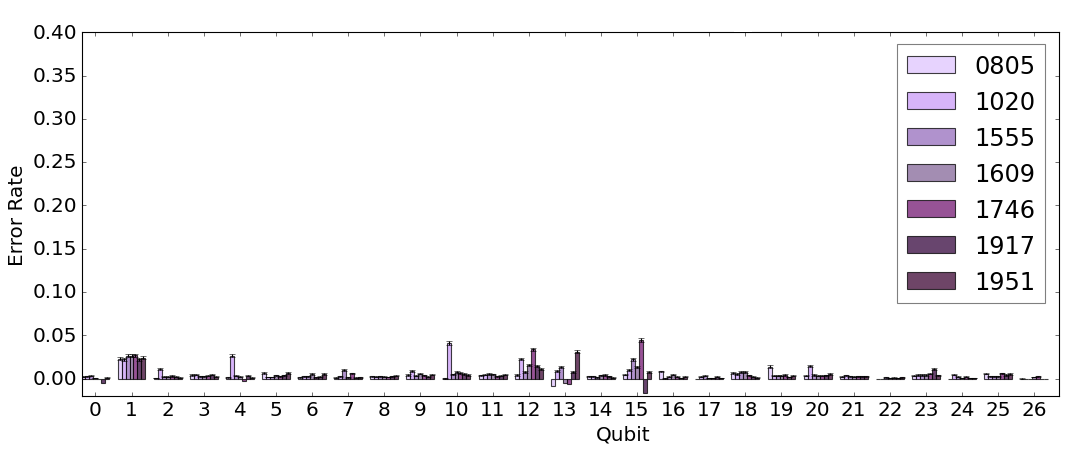}
    \caption{Depolarizing error rate for the $X$ gate on \toronto using the EDC methodology.}
    \label{fig:xcompare}
\end{figure}

We next report device characteristics across the time period these experiments ran. IBM Q devices are periodically calibrated, and this calibration includes sampling for measurement error approximately every hour. The results of these tests inform the calculation of the discriminator plane that distinguishes a measurement result of 0 from a result of 1 \cite{McKay_2017}. Figures \ref{fig:p0compare}, \ref{fig:p1compare}, and \ref{fig:xcompare} show 
results from EDC readout error analysis. 
\vspace{0.5in}
\vspace{0.5in}
\begin{figure}[htp]
    \centering
    \includegraphics[width=0.48\textwidth]{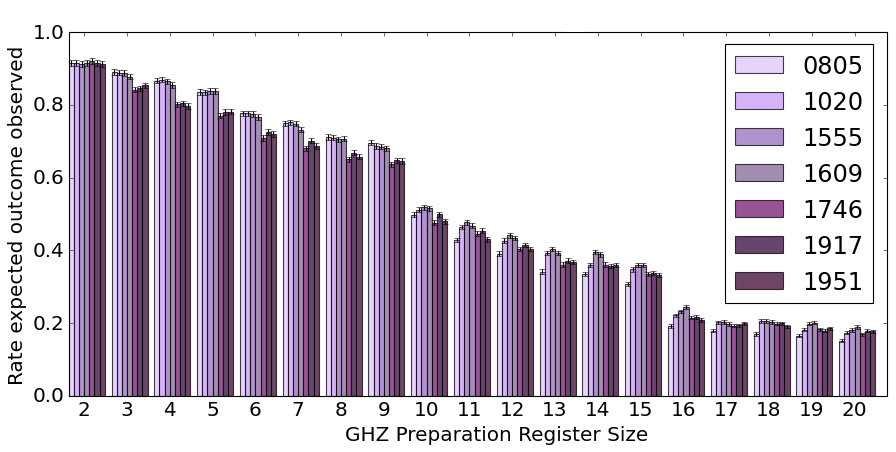}
    \caption{Rate expected outcomes were observed from BC GHZ-state preparation circuits executed on \toronto across a 12-hour period. The decay function of the best performing set (1555) is $1.112e^{-0.0834x}$ with $R^2=0.989$ for register size $x$ \cite{wolframalpha}.}
    \label{fig:bareGHZrates2-14}
\end{figure}
\vspace{1in}
\begin{figure}[hbp]
    \centering
    \includegraphics[width=0.48\textwidth]{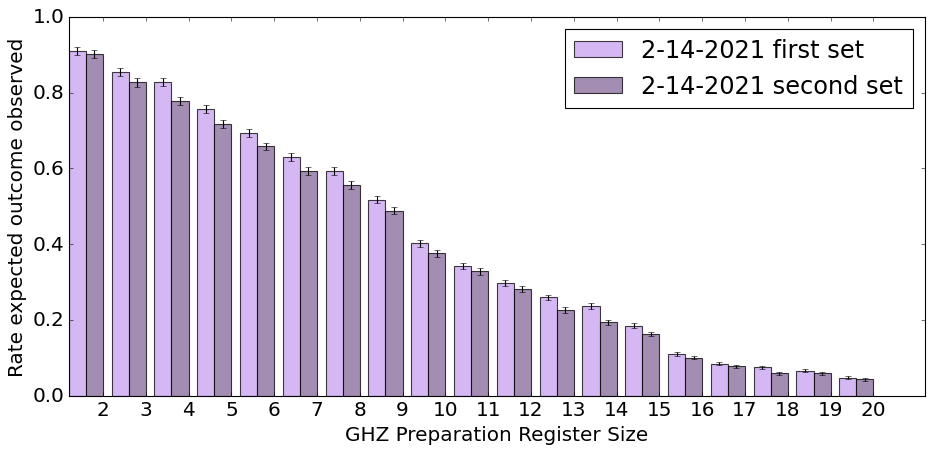}
    \caption{Rate expected outcomes were observed from RC GHZ-state preparation circuits executed on \toronto across a 12-hour period. Decay function of the best performing set (first set) is $1.148e^{-0.12x}$ with $R^2=0.9855$ for register size $x$ \cite{wolframalpha}. 
    }
    \label{fig:RCGHZrates2-14}
\end{figure}
\begin{figure}[htp]
    \centering
    \includegraphics[width=0.48\textwidth]{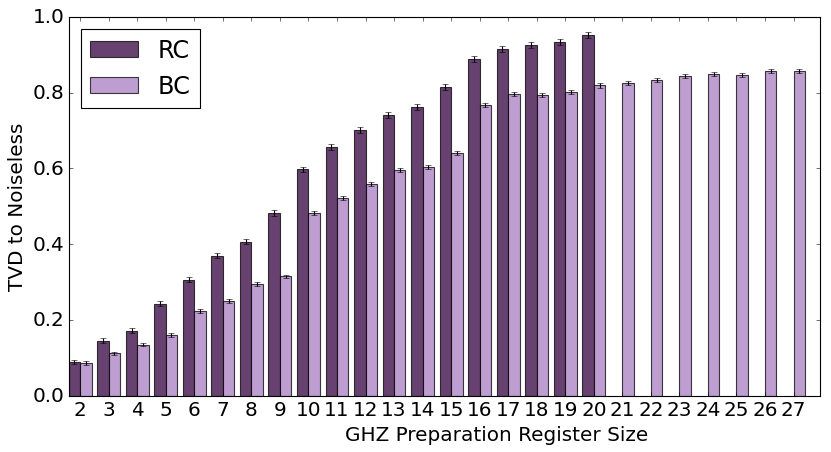}
    \caption{TVD between experiment results and noiseless GHZ-state preparations. RC circuits are limited to the first 20 qubits. The RC TVD increases with register size $x$ as $0.2418e^{0.0801x}$ with $R^2=0.978$ and the BC TVD increases as $0.2625e^{0.053x}$ with $R^2=0.963$ \cite{wolframalpha}.}
    \label{fig:rcbctvds}
\end{figure}
\par
For our selected RC and BC GHZ-state preparation circuits, we plot the TVD between the experimental results and noiseless results, i.e.~ideal outcomes, in Fig.~\ref{fig:rcbctvds}. For the noiseless case of GHZ-state preparations, we use an equal split of $N_s/2$ counts in state $\ket{0_0,...,0_n}$ and $N_s/2$ counts in state $\ket{1_0,...,1_n}$ for $N_s$ total shots and $n$ qubits. From these results, we see that the BC circuits are closer to the ideal outcomes than the RC circuits because their TVD remains closer to zero. This is likely because injecting twirling gates in the GHZ circuits can lead to a dramatic increase in the total gate count, which in this instance is most likely increasing the overall error rate of the circuits. The randomized compiling protocol compiles the twirling gates with neighboring single-qubit gates, but in the case of GHZ-state preparation all circuits consist only of two-qubit gates which are all twirled around. This is corroborated by the close agreement of RC and BC TVDs for the smallest GHZ states, when the total number of added twirling gates is lowest compared to the total gate count of the uncompiled circuit. 
\par
\begin{figure}[htp]
    \centering
    \includegraphics[width=0.48\textwidth]{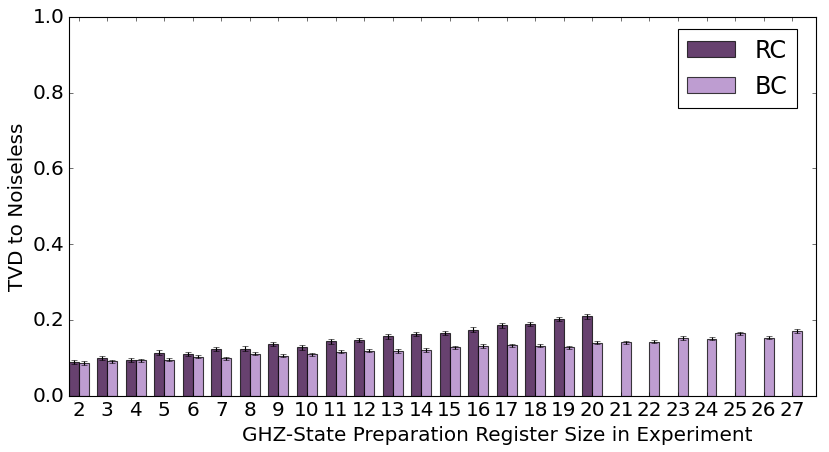}
    \caption{TVD between experiment results and noiseless GHZ results. For each size of GHZ-state preparation, the results are trimmed to the first two qubits. The RC TVD increases with register size $x$ as $0.0906e^{0.0445x}$ with $R^2=0.999$ and the BC TVD increases as $0.0889e^{0.024x}$ with $R^2=0.999$ \cite{wolframalpha}.}
    \label{fig:BCRCGHZbell}
\end{figure}

In Fig.~\ref{fig:BCRCGHZbell} we show the TVD between experiment results and noiseless GHZ-state preparations trimmed to only qubits 0 and 1 such that the full bitstring of each state observed is classified by the first two bits. We show these results to address the probabilistic decrease in observing a fully all-zero or all-one state from the largest register sizes. It becomes more likely that at least one bit of the bitstring outcome is flipped due to an error as the measured register size increases. The TVD of these results is reduced over the TVDs shown in Fig.~\ref{fig:rcbctvds}, which may also suggest that larger qubit registers correlate with higher error rates in experiment. However, the TVD increases steadily for larger sizes of GHZ-state preparation, which may capture the effects from decoherence on the first two qubits which idle while \cnot gates are performed on the other qubits of the register. 
\subsection{Characterization Results}
\label{subsec:protocolresults}
We report the characterization results of GST, KNR, and EDC of our experiments. 
\subsubsection{Gate Set Tomography}
\label{subsubsec:gstchar}
\par
Gate set tomography provides a detailed picture of the characterization of a defined gate set on a selected qubit subspace. We executed GST on qubits 0 and 1 of \toronto and obtain estimates of SPAM operators and the gates \{$R_X(\frac{\pi}{2})$,$R_Y(\frac{\pi}{2})$,$R_Z(\frac{\pi}{2})$,$I$,\textsc{cnot}\}, as well as estimates of the model fit and metrics of gate performance such as process fidelity. We calculate the completely-positive trace-preserving (CPTP) map that best fits the GST experiment data \cite{pygsti,nielsen2020probing}. The CPTP estimate results are about 45 standard deviations away for the shortest circuits of length 1 and 2 gates and about 250 standard deviations away from a Markovian gate set for the longest circuits of length 32 gates. This indicates the presence of non-Markovian noise, especially for longer gate sequences.
\par
\begin{figure}
    \centering
    \includegraphics[width=0.4\textwidth]{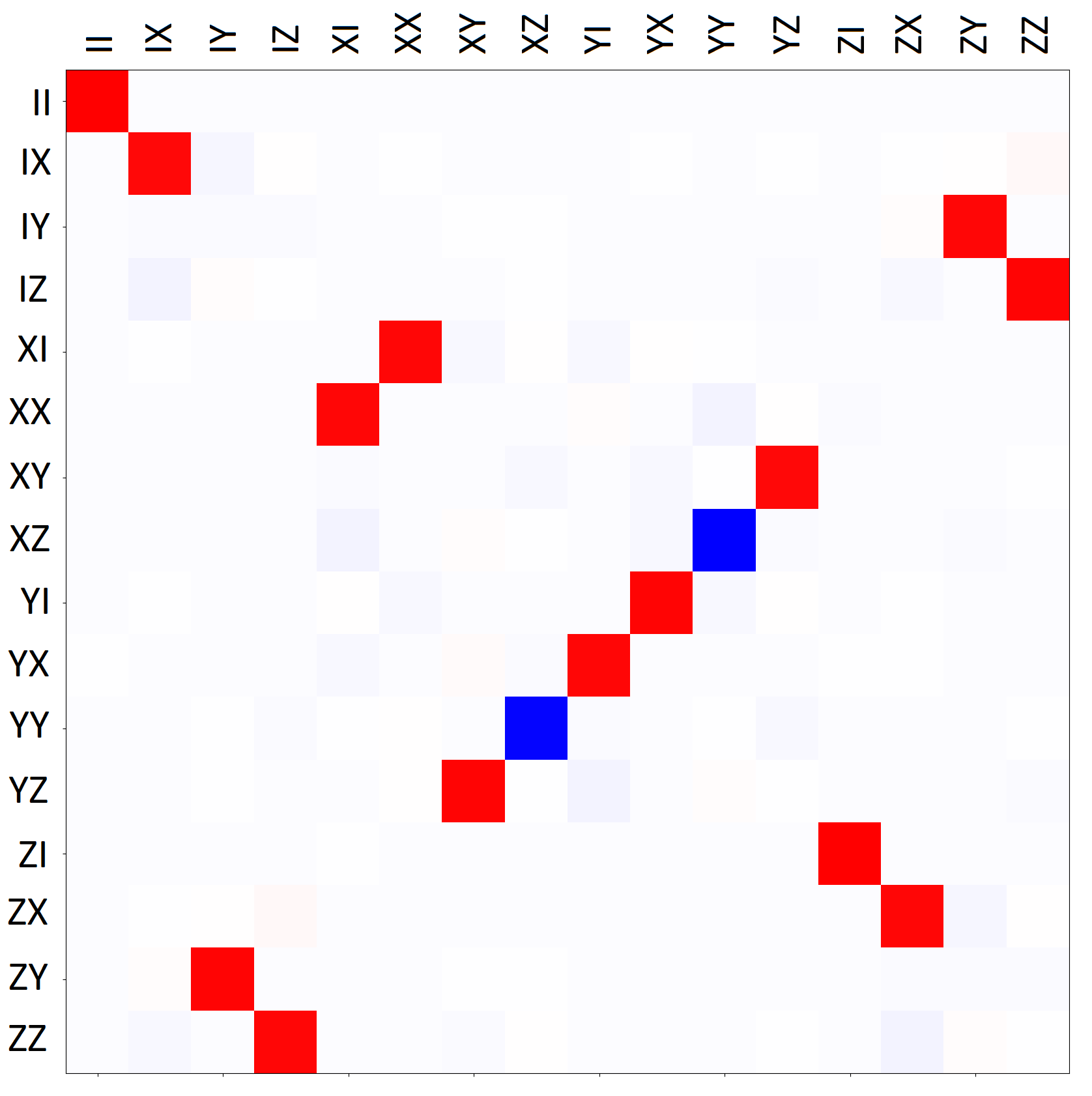}
    \caption{The Pauli transfer matrix estimated for \cnot on qubits 0 and 1 from GST \cite{pygsti}. The color scale ranges from red for values close to 1 and blue for values close to -1.}
    \label{fig:ptmcnot}
\end{figure}

In Fig.~\ref{fig:ptmcnot} we show the GST estimate of the Pauli Transfer Matrix (PTM) for the \cnot gate. The PTM represents the implementation of the operator in experiment. The ideal PTM consists of a single value of 1 or -1 in each row and column--noise manifests in the PTM in the non-zero terms which are lightly shaded in Fig.~\ref{fig:ptmcnot}. 
\par
\begin{figure}
    \centering
    \includegraphics[width=0.4\textwidth]{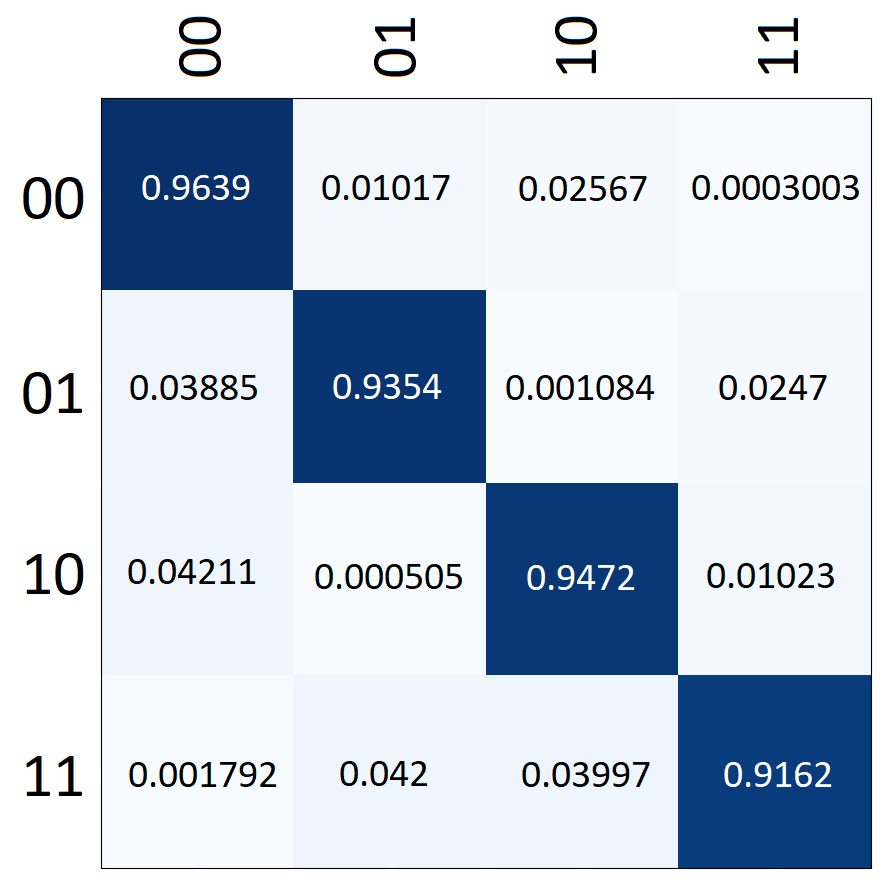}
    \caption{Readout matrix representing results from GST SPAM estimates.}
    \label{fig:gstreadoutmat}
\end{figure}

Figure \ref{fig:gstreadoutmat} shows the SPAM estimates from the GST model. We present the matrix of values which represent the probabilities of observing versus preparing each two-qubit state. The highest error rates are observed in the 11 state and the lowest are observed in the 00 state. This readout model inherently accounts for correlations in the two qubits by separately estimating the error on each two-qubit state.  
\par
\begin{figure}
    \centering
    \includegraphics[width=0.48\textwidth]{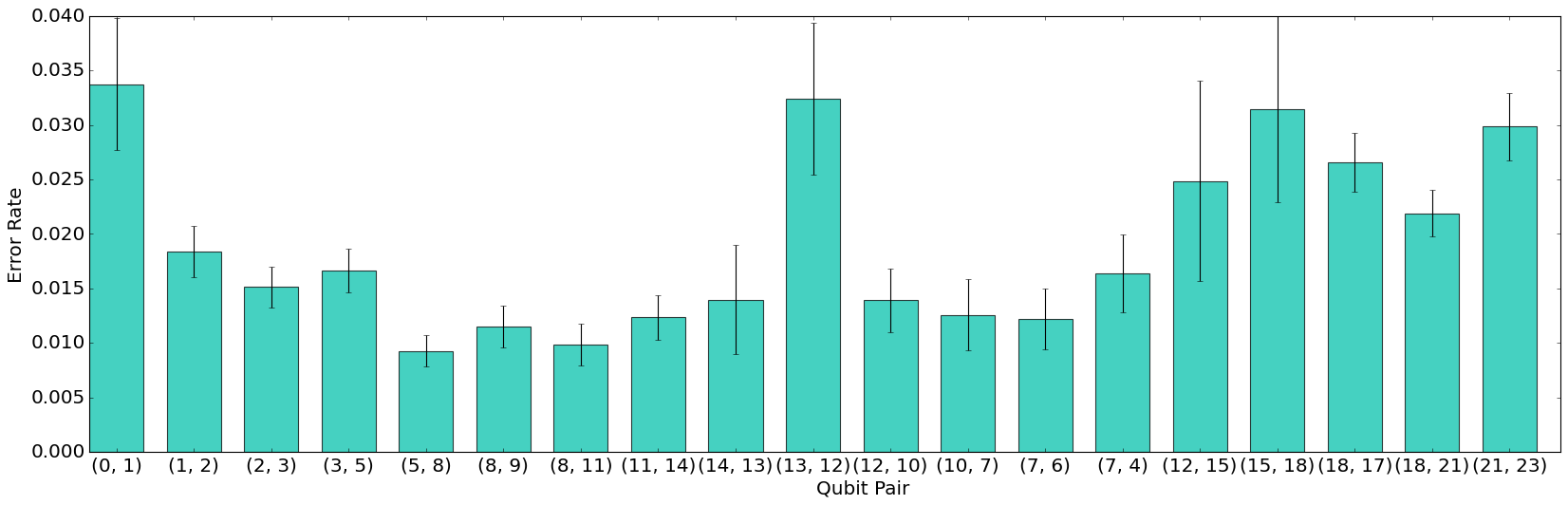}
    \caption{Total error rates for each \cnot characterized by KNR. Values represent the sum of all error types measured by the protocol.}
    \label{fig:knrcnoterror}
\end{figure}

From KNR, we obtain the full stochastic Pauli channel estimated for the specified cycle. In Fig.~\ref{fig:knrcnoterror} we show the error rates estimated using KNR where we have defined each cycle as a \cnot gate operating on a coupling on \texttt{toronto}. Some error types are indistinguishable in the KNR protocol for certain gates because the errors operate in the same way on the cycle of interest. For example, a \cnot gate cycle KNR result conflates $IY$ and $ZY$ Pauli errors. To construct our noise model, we preferentially select weight-one errors (any two-qubit Pauli operator that has an $I$ operator) where possible under the assumption that weight-one errors are more likely and assign the reported error probability to that error type. For indistinguishable weight-two errors there is no guiding principle for which error is more likely, so we select the first reported error of the two. Because the errors are lexicographically ordered there is a slight bias towards $X$-type errors, but we expect this to have little to no effect on the final results of the noise model since these error types are indistinguishable in practice.
\par
\begin{figure}
    \centering
    \includegraphics[width=0.48\textwidth]{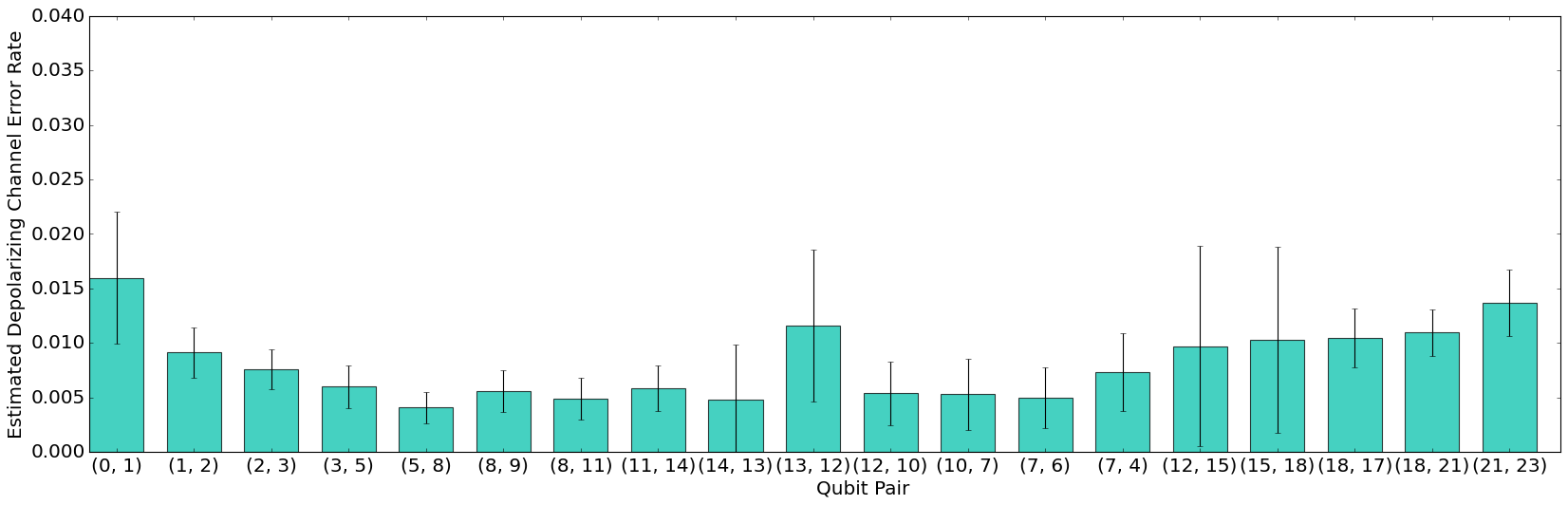}
    \caption{Estimate of the depolarizing component of the noise channels estimated by KNR.}
    \label{fig:knrdepolrates}
\end{figure}

In Fig.~\ref{fig:knrdepolrates} we show the estimated portion of these errors which could be modeled as a depolarizing channel. To estimate this parameter, we consider a single-qubit depolarizing channel like the one defined in Eq.~\ref{eq:isodepol} except that the parameter $p$ is allowed to vary per error gate operator ($X$, $Y$, or $Z$). We sum together the single-qubit (weight-one) error rates per qubit provided by the KNR estimate and average the two estimates together. This is an approximation of a depolarizing parameter that could describe the noise in a two-qubit gate as the channel $\epsilon^{\textrm{DP}}_{j,k} = \epsilon^{\textrm{DP}}_{j} \otimes \epsilon^{\textrm{DP}}_{k}$ in the same way EDC depolarizing estimates are defined. Several degrees of freedom that are estimated by KNR are ignored in this approximation but it is a useful comparison to the EDC-estimated depolarizing rates. 
\subsubsection{Empirical Direct Characterization}
\label{subsubsec:edcchar}
\par
\begin{figure}
    \centering
    \includegraphics[width=0.48\textwidth]{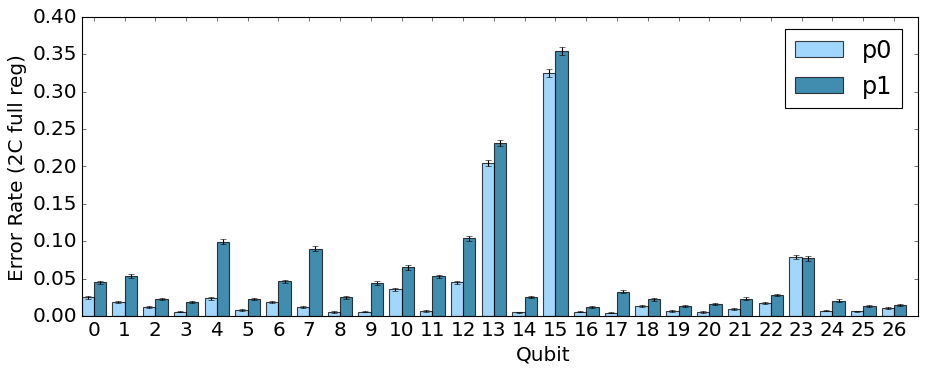}
    \caption{Readout error rates for \toronto estimated by EDC.}
    \label{fig:edcreadout}
\end{figure}

In Fig.~\ref{fig:edcreadout} we show estimated readout error rates for measurements of state zero and state one. The $p_0$ parameter is the rate of error in readout when state zero was the expected outcome; similarly the $p_1$ parameter is the rate of error in readout when state one was the expected outcome. These parameters are estimated from results of a blank measurement circuit and a circuit with a single $X$ gate applied to every qubit in parallel. We also test other methods of readout parameter estimation using one $X$ gate operation per qubit per circuit and adding circuits which use two $X$ gates to solve for error rates on $X$ such that the error rate $p_1$ is corrected for the error of applying $X$. However, this method provided the best performance in our tests which are shown in detail in Section \ref{subsec:comparisons}. These readout error estimates are also used in the KNR noise model, as this method for estimating readout is the same approach used in True-Q.
\par
The readout error rates indicate spatial variability across the qubit register, as well as a consistent asymmetry between states zero and one. In particular, most qubits have a higher error rate in readout of state one. Additionally, most qubits have under 5\% error rates, but qubits 13 and 15 are significant outliers with around 30\% error.
\par
\begin{figure}[htp]
    \centering
    \includegraphics[width=0.48\textwidth]{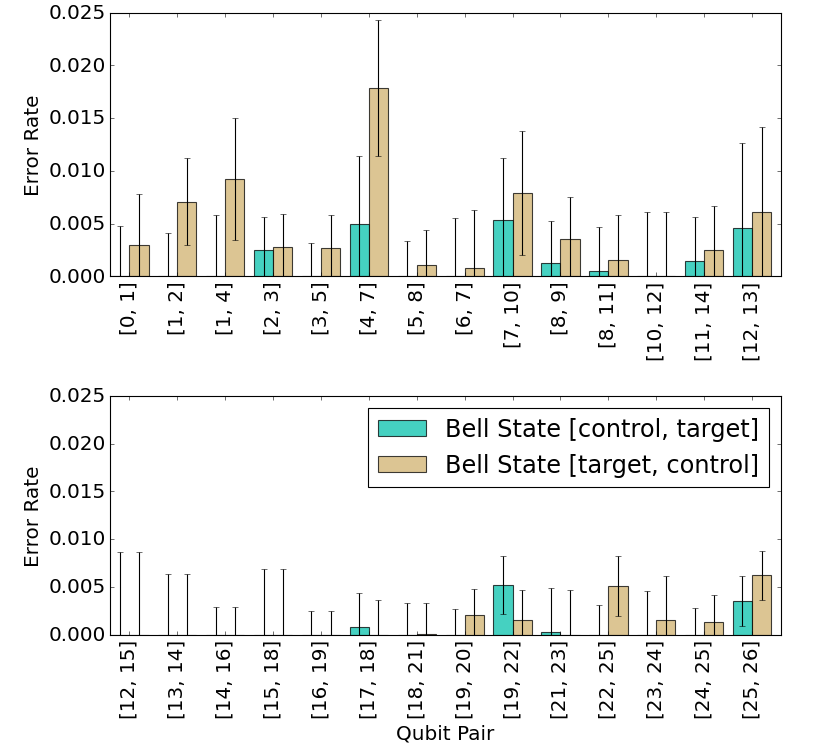}
    \caption{Error rates of \cnot for \toronto estimated by EDC. We observe spatial variability across the register and asymmetry in control versus target qubits in each coupling.}
    \label{fig:edccnotrates}
\end{figure}

Using EDC we calculate the depolarizing parameter which best fits Bell-state preparation circuit outcomes for Bell circuits executed on each qubit pair of the layout of \toronto shown in Fig.~\ref{fig:Tlayout}. In Fig.~\ref{fig:edccnotrates} we show these parameters which are evaluated for \cnot gates applied with both configurations of control and target qubits. The error bars represent the upper limit of the error from the least squares calculation. These error rates are calculated using the readout error rates from Fig.~\ref{fig:edcreadout}. 
\par
The results from EDC for depolarizing error rates show lower error rates and more spatial variability than the depolarizing estimates derived from KNR. The estimated error from EDC is frequently lower than the KNR depolarizing estimate, and the relative noise of the qubit couplings among each protocol estimate does not generally agree. The differences between the two estimates may largely be attributed to the use of RC in the KNR estimates, and this comparison provides a numeric estimate of the effect of RC on the observed error rates.
\subsection{Comparative Analysis of Characterization}
\label{subsec:comparisons}
\par
\begin{figure}[htp]
    \centering
    \includegraphics[width=0.48\textwidth]{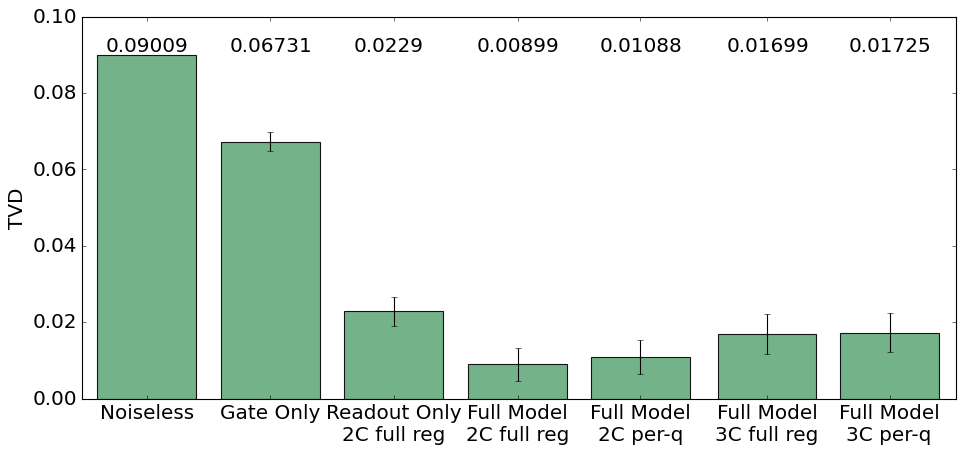}
    \caption{TVD to experiment of composite noise models constructed from error rates estimated using KNR. Error bars are calculated as the standard deviation across 100 trials of the Bell-state preparation circuit distributions. The full model with readout error based on just two circuits provided results closest to experiment as measured by TVD.}
    \label{fig:knrbell}
\end{figure}

In Fig.~\ref{fig:knrbell} we show the results of simulating a Bell-state preparation circuit using noise models derived from KNR results. We use the gate noise estimates from KNR shown in Fig.~\ref{fig:knrcnoterror}. The ``Gate Only'' noise model consists of just these error rates. We then add to this gate model four different methods of readout error. The readout error estimates are derived from using a single blank measurement circuit, a circuit with a single $X$ gate applied per qubit, and a circuit with two $X$ gates applied per qubit. The ``2C full register'' readout model uses the first two of these circuits to estimate readout. The ``3C full register'' readout model uses all three of these circuits to estimate readout. We can apply these $X$ and $XX$ gates once per qubit per circuit such that we have as many circuits as qubits. The motivation of this approach is to take any correlations between simultaneous operators into account. This approach is used in the ``2C per qubit'' and ``3C per qubit'' models.
\par
\begin{figure}[htp]
    \centering
    \includegraphics[width=0.48\textwidth]{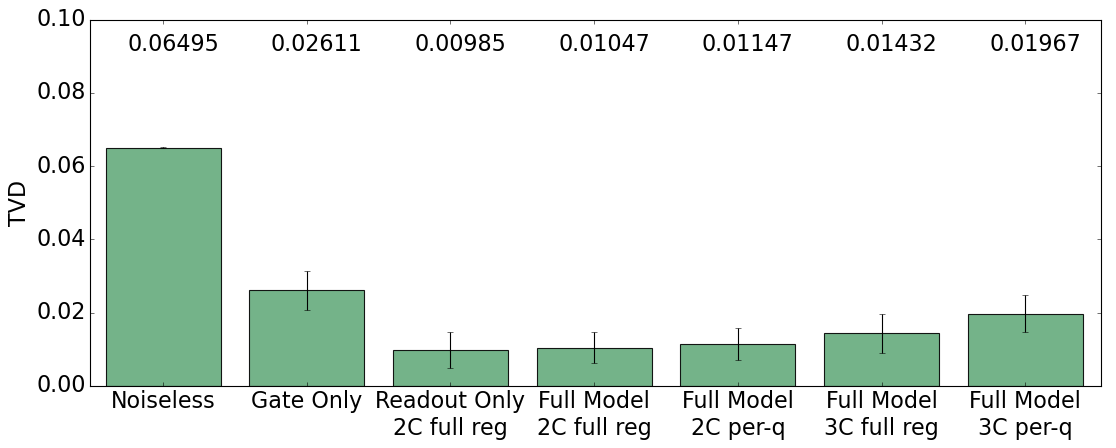}
    \caption{TVD to experiment of composite noise models constructed from error rates estimated using EDC. Error bars are calculated as the standard deviation across 100 trials of the Bell-state preparation circuit distributions. Although the readout only model provided results closest to experiment as measured by TVD, it is within error bars of the full model which is likely because the estimates for depolarizing noise in the gates were relatively low.}
    \label{fig:edcbell}
\end{figure}

In Fig.~\ref{fig:edcbell} we show these results for EDC models with the same set of four readout models. We use the gate noise estimates from Fig.~\ref{fig:edccnotrates}. For both the KNR and EDC results, we compare the performance of these noise models to the TVD between the experiment results and a noiseless Bell-state preparation, which is an exactly equal split between the 00 and 11 states, just as we defined the noiseless GHZ state.
\vspace{0.5in}
\par
\begin{figure}[htp]
    \centering
    \includegraphics[width=0.48\textwidth]{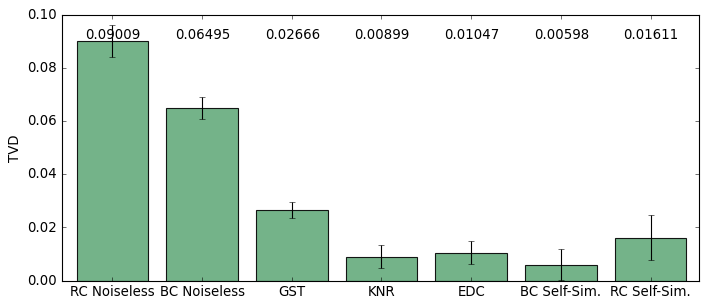}
    \caption{TVD between experiment and noisy simulation of the Bell state on qubits 0 and 1 using noise models constructed from GST, KNR, and EDC protocols. Error bars represent the standard deviation across 100 trials of Bell state simulation distributions. The error for the noiseless and self-simulation cases is calculated as the error propagation in TVD from the distributions.}
    \label{fig:fullbellcompare}
\end{figure}

In Fig.~\ref{fig:fullbellcompare} we show the results of simulating a Bell-state preparation circuit using the best noise model from EDC, KNR, and GST. We again compare to the TVD of the noiseless Bell-state preparation of equal counts of state 00 and state 11. We also compare to ``self-simulated'' cases, which is the TVD between the targeted Bell circuit results from experiment and another data set of Bell circuit results executed on \texttt{toronto}. The self-simulation examples indicate a potential best-case simulation of \toronto simulating itself by generating additional data sets. 
\par
The difference in performance of RC circuits and BC circuits is highlighted in the noiseless results. The lower TVD between the noiseless Bell results and the BC Bell results indicates that the BC Bell results are closer to ideal than the RC circuits. The KNR TVD is calculated to the RC Bell data, and the GST and EDC TVD is calculated to the BC Bell data. The noise model with the closest fit to experiment was KNR, although EDC is within error of KNR.  In the RC self-simulation case, the TVD result likely indicates the effects of drift, since the additional data set used for comparison was taken several hours later. For BC Bell circuits, the additional data set is taken from the same job. 
\subsection{GHZ Benchmark Results}
\label{subsec:ghzresults}
\par
\begin{figure}[htp]
    \centering
    \includegraphics[width=0.48\textwidth]{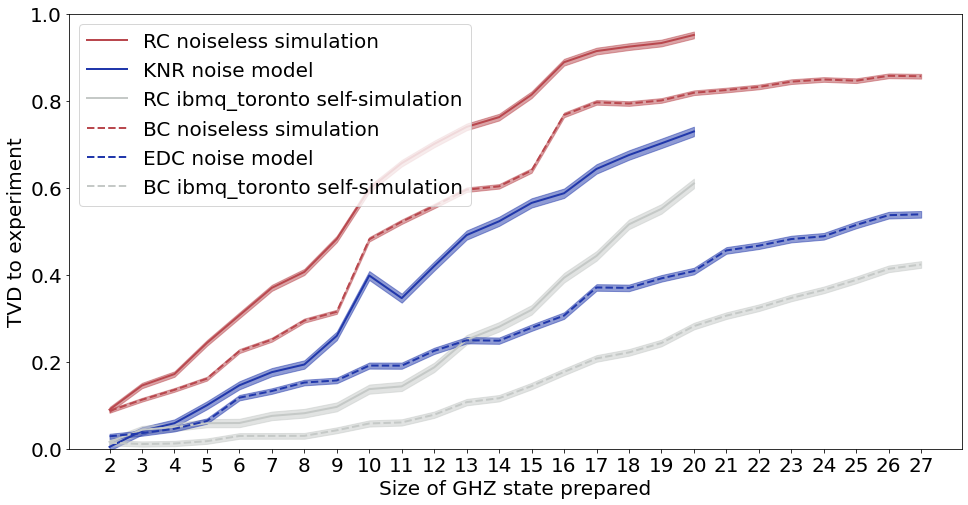}
    \caption{Total variation distance between experiment results and simulation results. Solid lines indicate TVDs calculated with randomly compiled results; dashed lines indicate TVDs with bare circuits. TVD for the noiseless case is calculated between experiment and results exactly split between an all-zero state and an all-one state. TVD for the ``self-simulation'' case is calculated between experiment and GHZ results executed on \toronto during the same 12-hour period.}
    \label{fig:edcknrghzsims}
\end{figure}

We evaluate the performance of noise models built using KNR and EDC methods in simulating GHZ-state preparation circuits. We calculate the TVD between our noisy circuit simulation outcomes and the circuit outcomes in experiment from \toronto and show these results in Fig.~\ref{fig:edcknrghzsims}. We compare these results to the TVD calculated between our selected RC and BC GHZ-state preparation circuit results and noiseless GHZ-state preparation results, for which we use an equal split between the all-zero and all-one states. We also compare to the TVD calculated between the selected RC and BC GHZ-state preparation circuit results and an additional data set of the same circuits run on \toronto during the same time frame. 
\par
We find that the EDC noise model comes closest to accurately simulating the results of the GHZ circuits in experiment. The EDC noisy simulations are both closer to the self-simulated (best case) results and farther from the noiseless (worst case) results than the KNR noisy simulation results are to the respective RC GHZ results. At a size of 20 qubits, the EDC noise model simulation is about 0.4 lower TVD than noiseless, whereas the KNR noisy simulation is about 0.2 lower TVD than noiseless. 
For GHZ circuits of size 2 and 3 qubits, KNR simulation TVD is lower than the self-simulated TVD, but reaches a maximum distance away from self-simulated of 0.26 at 10 qubits, whereas EDC simulation TVD reaches a maximum distance from self-simulated of only 0.16 at 17 qubits.
\par
These results indicate that the EDC noise model provides a closer description of the noise present in \toronto than the KNR noise model.
\subsection{Bernstein-Vazirani Benchmark Results}
\label{subsec:bvresults}
\par
\begin{figure}
    \centering
    \includegraphics[width=0.48\textwidth]{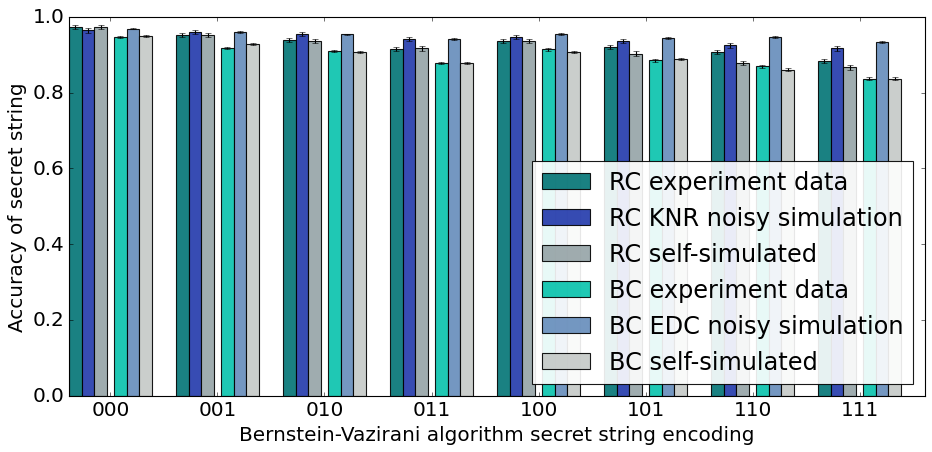}
    \caption{Bernstein-Vazirani results from experiment and noisy simulation. The accuracy of the secret string is the number of times the correct encoding is measured out of all measurements of the BV circuit. Higher accuracy indicates less noise present in the quantum circuit.}
    \label{fig:bvcompare}
\end{figure}

In Fig.~\ref{fig:bvcompare} we show the performance of our EDC and KNR noise models on the Bernstein-Vazirani algorithm. We compare these to the performance of the targeted experiment. In the case of EDC, this is the set of BV circuits which were executed in the same job as the characterization experiments and GHZ circuits. In the case of KNR, this is the set of BV circuits which were executed closest in time following the KNR characterization experiments. The EDC noise model used to model GHZ circuits is sufficient to model BV circuits because it consists of the same components. We use the KNR protocol to characterize cycles which define BV circuits and construct a noise model from these results as detailed in Section \ref{subsec:charexpstq}. 
\par
We compare the results of noisy BV circuit simulations to a ``self-simulating'' experiment, which for EDC is a set of BC BV circuits from a job about 15 minutes later and for KNR is a set of RC BV circuits executed a few hours later. The time differences are due to the amount of experiments which happened in between. In particular the short time between EDC circuit trials is a result of running the EDC circuit set multiple times to track error over time as in Figs.~\ref{fig:p0compare} and \ref{fig:p1compare}. BV secret strings should be returned by 100\% of the results in the absence of noise, and therefore the noiseless case of BV circuit implementation returns an accuracy of 1.
\par
The randomly compiled BV circuits performed better than the bare BV circuits for every secret string encoding. The KNR noise model also performed better than the EDC noise model for every secret string, coming closer to the BV circuit results from \toronto though both noise models were outside of error bars in every bitstring example. For encoded strings 101 and 110, the KNR noise model performed about as well as the self-simulated results.
\par
Additionally, for both noise models the difference in accuracy between the noisy simulation and experiment increases with the number of \cnot gates. 
It might indicate that there are additional noise sources present in \cnot that are not accounted for in either the KNR model or the EDC model. 
\subsection{Classical Resources Usage}
\label{subsec:classicalresources}
\par
We evaluate classical computation expense for these methods in creating experiments, analyzing the data, and performing classical simulations of quantum circuits. We report these in Table \ref{tab:classicalresources} measured by the amount of time taken to complete the computation. In general, the computational intensity of creating experiments is negligible. Although creating a GST experiment set can be computationally intensive, we are using a pre-built experiment set in pyGSTi. Analyzing the data of KNR and EDC requires a trivial amount of time to calculate on a basic laptop, but calculating the results of GST is computationally intensive as the algorithm for the GST analysis optimizes a model to best fit the experiment data \cite{nielsen2020probing}. 
\par
\begin{figure}[htp]
    \centering
    \includegraphics[width=0.48\textwidth]{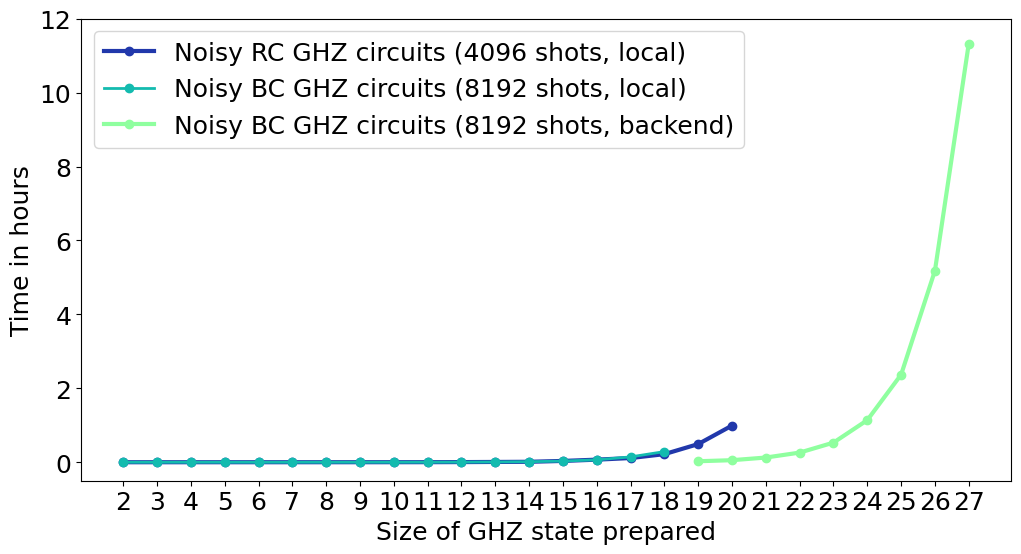}
    \caption{Time taken for noisy GHZ simulation as a function of qubit count. Noisy BC GHZ circuit simulations were switched from the local laptop to the IBM Q simulator backend for 19+ qubits due to the computational intensity of the largest GHZ circuit simulations. The IBM Q simulation backend has a limit of 32 qubits \cite{ibmqsuite}.}
    \label{fig:simulationtiming}
\end{figure}

Classical simulations of noisy quantum circuits are notoriously intensive, and we present a detailed report of their performance in Fig.~\ref{fig:simulationtiming}. The classical computational expense of simulating quantum circuits grows exponentially in the size of the qubit count, and this trend is demonstrated in all of our noisy GHZ-state preparation circuit simulations. 
\begin{table}[htp]
    \centering
    \caption{Estimates of classical resources used in our selected protocols. The ``local'' simulations were computed on a laptop with 16 GB RAM and Intel Core i7 processor. The ``backend'' simulations were sent as jobs to the IBM Q \texttt{ibmq\_qasm\_simulator}, a dedicated quantum circuit simulator backend which is available through the IBM Q suite \cite{ibmqsuite}. While the GST classical calculation is computationally expensive, it may be parallelized on multiple processors to achieve speedup.}
    \begin{tabular}{|l|l|l|}
        \hline
        \textbf{Method} & \textbf{Details} & \textbf{Time Taken} \\
        \hline
        GST & \makecell[l]{Calculate results} & \makecell[l]{70.13 hours} \\ \hline
        KNR & \makecell[l]{Simulate noisy RC GHZ circuits \\ (2-20 qubits, local)} & \makecell[l]{1.95 hours}  \\ \hline
        EDC & \makecell[l]{Simulate noisy BC GHZ circuits \\ (2-18 qubits, local)} & \makecell[l]{33.53 minutes} \\
        \hline
        EDC & \makecell[l]{Simulate noisy BC GHZ circuits \\ (19-27 qubits, backend)} & \makecell[l]{21.02 hours} \\ \hline
        KNR & \makecell[l]{Simulate noisy RC BV circuits \\ (8 four-qubit circuits compiled \\ into 32 circuits each)} & \makecell[l]{1 minute} \\ \hline
        EDC & \makecell[l]{Simulate noisy BC BV circuits \\ (8 four-qubit circuits)} & \makecell[l]{1 minute} \\ \hline
    \end{tabular}
    \label{tab:classicalresources}
\end{table}

\section{Conclusion}
\label{sec:conc}
\par
We have implemented several different characterization methods in experiment on quantum computers. We demonstrate the effectiveness of each method in estimating device parameters and the accuracy of the resulting noise model in describing the QPU outcomes. We show that EDC, a highly efficient approach which yields a coarse-grained noise model, offers competitive accuracy with other state-of-the-art methods in tests on standard quantum algorithms. 
\par
Our results demonstrate that the best characterization method depends on the application. The structure, components and size of the circuit to be characterized all play a role in choosing characterization approaches. Furthermore, our results show that an increase in experiment count and consequently information gain does not correlate with higher accuracy in noise descriptions used to simulate a QPU.
\par
Our BV circuit results can be further applied as a benchmark of QPU performance. While BV algorithm circuits are unlikely to represent a broad indication of QPU performance, they share some hallmarks of benchmark tests with other commonly used benchmarks. For instance, BV circuits represent a practical calculation that a QPU might be used to solve. They are easily extensible and include several features of a typical quantum computation, such as single- and two-qubit gates, entanglement, superposition, and measurement. Furthermore, the expected outcome of BV is a single bitstring, which means that any results that are not the expected bitstring are the result of errors. 
\par
In this way, KNR may provide an estimate of performance of a QPU on BV algorithm benchmarks. KNR had close agreement to experiment in the reported accuracy of the encoded secret string. Using noise models constructed by KNR in simulation of BV benchmark experiments could provide valuable insight into expected performance of a QPU on a relevant test.
\par
None of the models derived from the three methods were able to describe all noise present in the device. We tested all three on the Bell state and none of the results achieved zero TVD, although the TVD was low for KNR and EDC. In addition, the TVD results for GHZ circuits grew with the size of the circuit qubit register, indicating not only that there are additional soures of noise present in the system but that they might correlate with qubit count. In the GHZ example, some of this additional noise may come from decoherence of idling qubits as shown in Fig.~\ref{fig:BCRCGHZbell}. However, TVD sharply increases for larger circuits, approaching the maximum of one for the maximum register sizes (20-27 qubits), indicating that these experiments reach the limit of the capabilities of the QPU's ability to produce correct or predictable results.
\par
The best characterization method varied by test. EDC performed better in GHZ tests, and KNR performed better on BV benchmarks. The structure of the quantum circuit might be a factor in this difference. To improve EDC performance on BV it may be beneficial to re-characterize the components in the BV circuit context and define a set of tests to characterize more components of the circuit. To improve KNR performance on GHZ, we may be able to change the experiment setup in a way that is more optimized for the circuit and set of cycles, such as testing different combinations of \cnot parallelizations or sequence lengths.
\par
If the performance of RC circuits and KNR noise models are strongly correlated and the performance of BC circuits and EDC noise models are similarly correlated, it may be possible to test what characterization method is most useful for a particular circuit. We could test the circuit of interest or a subcircuit of it for which we know the expected outcome both with and without randomized compiling. If the RC circuits perform better than the BC circuits, that might indicate that KNR is the best noise model choice. Likewise, if the BC circuits perform better than RC circuits, EDC may be the better noise model.
\par
The main findings are generally corroborated among the protocols. For instance, the relative noise levels between asymmetric readout states and single- and two-qubit gates are similar among the protocols. However, the fidelity metrics and gate error levels are not always in agreement.
\par
Non-Markovian noise is present in the system and particularly in the \cnot gates. GST indicates this, as does the performance of RC circuits to a lesser extent. RC circuits should perform best at tailoring noise into stochastic Pauli channels in the presence of arbitrarily non-Markovian noise \cite{wallman2015noise}. Because the performance of KNR noise models steadily degrades for larger counts of \cnot gates, these stochastic channels evidently do not predict the QPU results and therefore the noise has not been tailored well.
\subsection{GST}
\par
The expense of GST is not prohibitive for a two-qubit example. However, it is large enough that over the time period that GST experiments are run it is possible that parameter drift comes into play which affects the accuracy of the best-fit model. Drift is more likely to impact GST results than KNR, or EDC primarily because GST requires the most experiments. To characterize a similar two-qubit example using KNR or EDC requires about 1/40th or less of the circuit count of GST. 
\par
However, the results of GST have more information to offer. In our experiments, GST confirms the presence of non-Markovian noise particularly prevalent in \cnot gates, which are also the noisiest gates in our tests. Non-Markovian noise impacts the accuracy of noise models built using GST, KNR, and EDC, and is likely to be a source of the additional error not accounted for in our best-fit models.
\subsection{KNR}
\par
In our GHZ circuit tests, the KNR noise model did not account for all the noise in the system. The KNR model achieved a poorer fit to RC GHZ experiment results than the EDC model fit to the BC GHZ experiment results as measured by TVD. Furthermore, the RC GHZ circuit results themselves have generally lower performance than uncompiled GHZ circuit results as measured by the rate of observation of the expected outcomes. As the circuit qubit register size increases, this performance worsens more quickly in the RC GHZ circuits than in the uncompiled GHZ circuits.
\par
On the other hand, the KNR noise model performs better than the EDC noise model on the BV circuit benchmark as measured by the agreement between simulation and experiment in accuracy reported by the algorithm. This was true for every encoded secret string. Additionally, the RC BV circuit results had better performance than the uncompiled BV circuits by this same measure for every encoded secret string. 
\par
These two aspects of the KNR protocol results--performance of the KNR model fit and performance of the RC application circuit--are likely correlated. In the GHZ example, the performance of the KNR model and the RC circuits was poor yet in the BV example, the performance of the KNR model and the RC circuits was good. One likely reason for this is that while the gate set of GHZ and BV is the same, the structure of GHZ circuits is very different from the structure of BV circuits. GHZ circuits are a chain of \cnot gates with no cyclical structure, no parallelized gates, one easy gate and virtually all hard gates. In contrast, BV circuits have a cyclical structure, many parallelized gates, many more easy gates than hard gates, and only a few hard gates that are applied right before or after easy gates which allows twirling gates to be compiled together with the easy gates. The way that hard gates are used in the circuit is likely a primary factor in the performance of RC and KNR noise models. The chain of unparallelized hard gates of GHZ circuits means that each hard gate becomes its own cycle and randomly compiling GHZ circuits results in a potentially large number of twirling gates inserted around each cycle. For the largest GHZ circuit examples we implemented, this can be up to an additional 300 single qubit easy gates inserted into the GHZ circuit. For the BV circuits, the additional gate count is no more than 8. This means that the potential for additional noise is much higher for the GHZ example than the BV example, leading to a commensurate degradation in performance of the application circuit. Likewise, this would have the effect of altering the noise channels measured by the KNR protocol such that the results of KNR may not be sufficiently descriptive of the application circuit to yield an accurate noise model.
\subsection{EDC}
The model of EDC is the simplest of all the methods and therefore provides the least detail of the underlying device or circuit characteristics. However, the EDC model provides a description of noise present on every tested component of the QPU and yields a noise model which performs best in simulating GHZ circuits as measured by TVD to experiment. It also requires the fewest experiments and scales only linearly in the characterized components, making it the most efficient approach to characterization.
\par
While the EDC model did not perform as well as the KNR model in simulating BV circuit results, it does not provide a noise model for the single-qubit gates present in the BV circuits. Developing a noise model for single-qubit gates using the EDC method may improve the accuracy of the EDC model in the BV example.
\par
In the BC GHZ circuit results, there are sharp increases in TVD between experiment and a noiseless GHZ when unusually noisy qubits are included in the circuit. These results demonstrate that it is worthwhile to avoid low-quality qubits. These results from \toronto would likely be improved using routing techniques to better handle highly noisy qubits \cite{Tannu_2019}.
\subsection{Computational Resources}
A central focus of our tests is scalability, namely how accuracy of characterization correlates with experiment count. Our results suggest that this is not a strong correlation. EDC has a low resource count but high accuracy in some of our tests. KNR had high accuracy in other tests but requires significantly more circuits than EDC. GST has the highest experiment count which yielded a suite of information about the 2-qubit system but did not perform well in the Bell state test.
\par
Classical computation resources needed to calculate GST are considerably higher than those needed for other protocols. These are not prohibitive in our example and can be reduced over our reported classical performance using parallelization and enhanced classical hardware. However, it is noteworthy that the classical computation expense of GST is a consideration in the overall experiment design, whereas for EDC and KNR the classical portion of the methodology is negligible.
\par
While the experiment count of KNR does not scale with qubit count, the total number of experiments needed to characterize all necessary cycles of the circuit of interest might still be high. Efficiency in experiments can be tuned in selecting the sequence lengths, number of sequence lengths, shot count, and circuit decompositions into cycles.

\section*{Acknowledgements}
{This research is supported by the Department of Energy Office of Science Early Career Research Program and used resources of the Oak Ridge Leadership Computing Facility, which is a DOE Office of Science User Facility supported under Contract DE-AC05-00OR22725.}
\bibliography{ref}
\bibliographystyle{plain} 

\appendix
\section{Bell State and GHZ States}
\label{app:bellghz}
\begin{figure}[htb]
    \centering
    \includegraphics[width=0.3\textwidth]{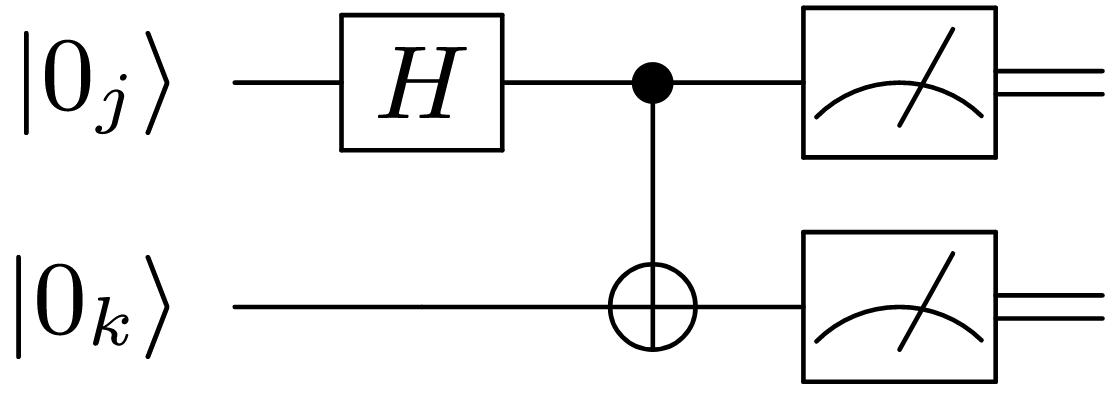}
    \caption{Bell state preparation circuit. }
    \label{fig:bellcirc}
\end{figure}

The Bell state is an equal superposition state of two entangled qubits and prepared using the circuit shown in Figure \ref{fig:bellcirc}. The $n$-qubit GHZ state is the extension of the Bell state to $n$ qubits and expressed as in Eq.~\ref{eq:ghzstate}. To prepare an $n$-qubit GHZ state, we add \cnot gates to the circuit shown in Fig.~\ref{fig:bellcirc} in a cascading ladder pattern such that each qubit of the GHZ state is a target of a \cnot gate but only the last qubit is not a control for a \cnot gate.
\begin{equation}
    \ket{\textrm{GHZ}(n)} = \frac{1}{\sqrt{2}}\left(\ket{0_1,0_2,...,0_n} + \ket{1_1,1_2,...,1_n}\right)
    \label{eq:ghzstate}
\end{equation}

\section{Bernstein-Vazirani Algorithm}
\label{app:bv}
\begin{figure}[hb]
    \centering
    \includegraphics[width=0.4\textwidth]{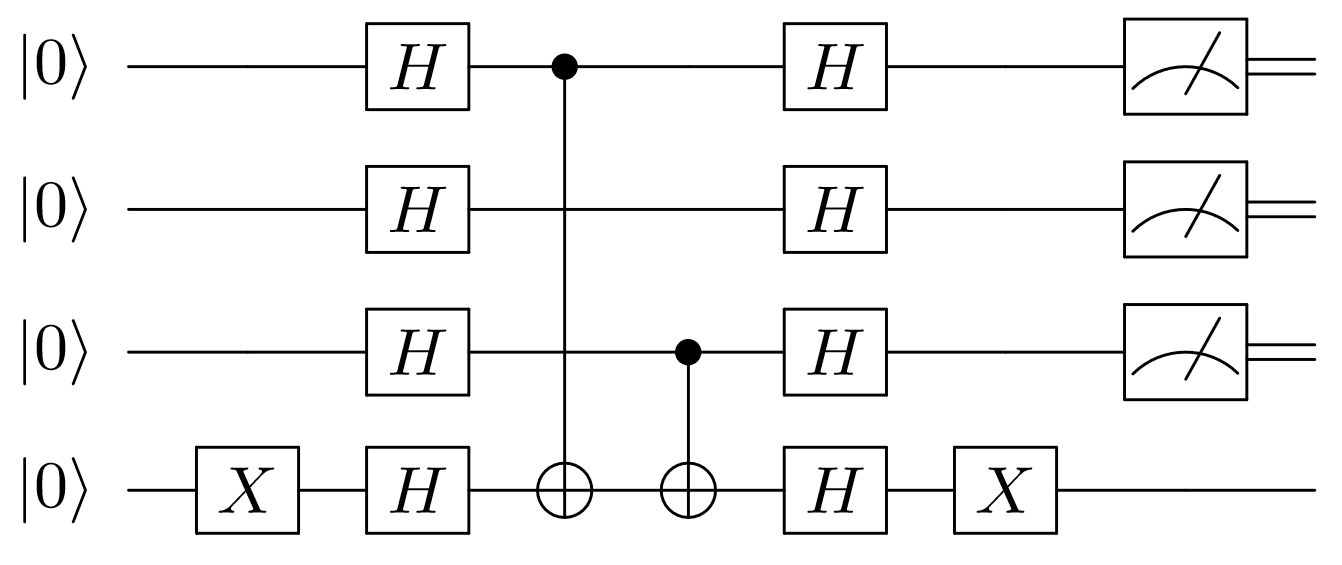}
    \caption{Implementation of the Bernstein-Vazirani algorithm shown for an encoded binary secret string of 101.}
    \label{fig:bvalgorithm}
\end{figure}

The Bernstein-Vazirani algorithm uses a black-box oracle to learn a secret string in a single query \cite{BValgorithm}. Figure \ref{fig:bvalgorithm} shows the quantum circuit implementation we use in simulation and experiment.
\end{document}